\documentclass[aps,prl,reprint,superscriptaddress,longbibliography]{revtex4-2}
\usepackage{amsmath,amssymb,bm,graphicx,booktabs,microtype,needspace}
\usepackage[T1]{fontenc}
\usepackage{lmodern}
\usepackage[hidelinks]{hyperref}
\hypersetup{pdftitle={Quantum Representation Selects Neutral Skyrmion Molecules},pdfauthor={Vishnu Jejjala}}
\makeatletter
\AtBeginDocument{\let\LS@rot\@undefined}
\makeatother
\renewcommand\sec[1]{\par\Needspace{3\baselineskip}\noindent\textit{#1.}\quad}

\begin{document}
\title{Quantum Representation Selects Neutral Skyrmion Molecules}
\author{Vishnu Jejjala}
\email{v.jejjala@wits.ac.za}
\affiliation{Mandelstam Institute for Theoretical Physics, School of Physics, University of the Witwatersrand, Johannesburg 2000, South Africa}
\affiliation{National Institute for Theoretical and Computational Sciences, South Africa}
\date{September 2026}
\begin{abstract}
Skyrmions experience a sideways gyroscopic force when they move.
Bound molecules can cancel this force; we call them gyroscopically neutral.
We show that the quantum states available on each site of a magnet select the composition needed for cancellation.
In two models of $SU(3)$ magnetism, this composition changes from a pair to a trimer.
Both models assign exactly the same energy to every configuration of their common classical order parameter, isolating the effect of the local quantum states.
Neutral molecules can also unwind when additional local states are allowed.
Mean field calculations nevertheless find both compositions bound and locally stable, with the neutral trimer surviving sampled interaction changes.
Comparing their motion reveals the changed cancellation rule, and prepared neutral molecules translate over short distances with little deformation.
\end{abstract}
\maketitle

A magnetic skyrmion is a pattern of local order that behaves as a particle.
Its motion produces a sideways gyroscopic force, the term responsible for skyrmion Hall deflection under an applied drive~\cite{Thiele,Stone}.
When skyrmions bind into a molecule, their gyroscopic forces can cancel~\cite{KomineasPapanicolaou,KomineasChiral,BarkerTretiakov,PanigrahySAF}.
We call this cancellation \emph{gyroscopic neutrality}.
The local quantum state helps determine the force: for an ordinary spin, its strength scales with the spin quantum number.
A magnet with several internal components offers more freedom because different windings can carry different gyroscopic weights.
Here, we show that this freedom changes the neutral molecular composition from a pair to a trimer, even at a fixed energy landscape for the ordered states.

We study a square lattice in two spatial dimensions with $SU(3)$ degrees of freedom on each site.
The local quantum representation, or multiplet, plays the role that the spin quantum number plays in an ordinary magnet: it specifies the available local states and their response to internal rotations.
We compare the eight state $(1,1)$ multiplet and the fifteen state $(1,2)$ multiplet.
Their dimensions follow from $d_{p,q}=(p+1)(q+1)(p+q+2)/2$.
An orbital picture makes these state counts concrete.
Fermions with three internal states carry a three state multiplet when an orbital holds one particle, and its conjugate when it holds two particles, identified by the missing state.
Combining one particle orbital with one or two hole orbitals, and projecting onto the selected irreducible multiplet, gives these eight or fifteen local states~\cite{GorshkovTwoOrbital,ScazzaOrbitalExchange,Supplemental}.
Our interactions are designed to isolate the representation effect; an implementation of the complete Hamiltonian remains to be developed.

The ordered states of both multiplets are described by the same \emph{flag}: two orthogonal directions $v,w$ in an internal complex space, with the third direction fixed by orthogonality.
Each direction can carry an integer winding, giving two species of skyrmion.
These states are the $SU(3)$ coherent states, the counterpart of a spin pointing in a definite direction; we call them \emph{flag states} below.
Earlier work established soliton pairing, multipolar textures, fractional skyrmion molecules, and interacting $SU(3)$ solitons~\cite{IvanovPairing,ZhangCP2,AkagiFractionalMolecules,UedaAkagiShannon,AmariFlagSolitons}.
The dependence of flag Berry phases on the local representation is also established~\cite{KimOh,AffleckBykovWamer,LajkoChains,BykovChains,WamerChains,TanizakiSulejmanpasic,OhmoriSeibergShao}, as is the distinction between neutral and charged pair motion in projective models~\cite{LeeJeonHan}.
Our contribution is a controlled realization in which the local multiplet selects \emph{which composition} is neutral while the energy of every flag texture remains exactly fixed.

\begin{figure*}[t]
\includegraphics[width=\textwidth]{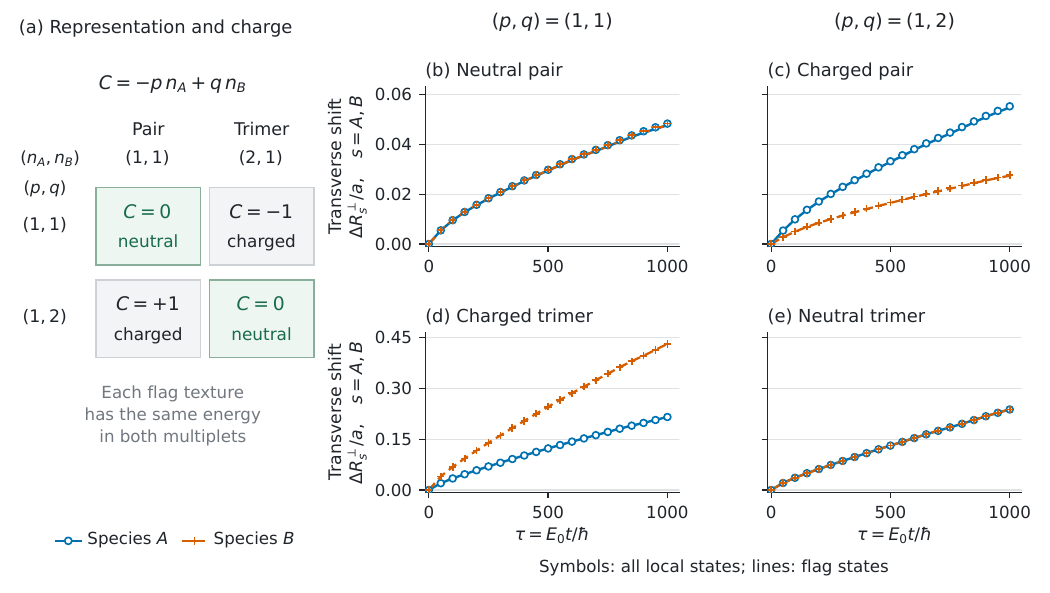}
\caption{Changing the local multiplet exchanges the neutral composition.
(a) The gyroscopic charges of species $A$ and $B$ are $-p$ and $+q$; the same flag texture has the same energy in either multiplet.
(b,c) A common stretched pair gives approximately equal species displacements in $(1,1)$ and a $2{:}1$ ratio in $(1,2)$.
(d,e) A common trimer preparation gives the reciprocal $1{:}2$ and $1{:}1$ responses.
Symbols let all local amplitudes evolve; lines keep each site in a flag state.
$\Delta R_s^\perp$ is the displacement of the center of species $s=A,B$, measured perpendicular to the preparation axis.
For full states, the winding centers use the highest and lowest eigenlines of the local $SU(3)$ moment~\cite{Supplemental}.
For a trimer, the $A$ center averages the two outer lobes shown in Figure~\ref{fig:binding}.
The pair and trimer rows use different displacement scales.}
\label{fig:selection}
\end{figure*}

\begin{figure*}[t]
\includegraphics[width=\textwidth]{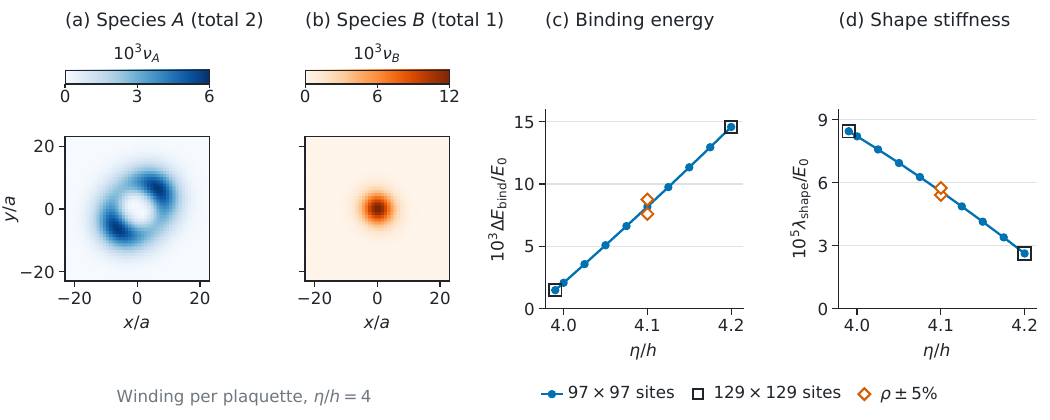}
\caption{The neutral trimer remains bound while becoming easier to deform.
(a,b) Winding per plaquette at $\eta/h=4$, with the sign of the core winding taken as positive: $\nu_A=-q_v$ and $\nu_B=-q_w$.
The complete sums are two and one; the color scales are independent.
Two $A$ lobes at $\pm(5.73,5.73)a$ flank the central $B$ lobe, forming a diagonal molecule.
(c) Binding energy $\Delta E_{\rm bind}=E_{A_2}+E_B-E_{A_2B}$ relative to separately optimized fragments.
(d) Shape stiffness $\lambda_{\rm shape}$, the lowest energy curvature for a change of shape, excluding translation and rotation from the plot.
Both remain positive across ten sampled attraction strengths.
Squares increase the lattice size; diamonds change the exchange stiffness $\rho$ by $\pm5\%$.}
\label{fig:binding}
\end{figure*}

\begin{figure}[t]
\includegraphics[width=\columnwidth]{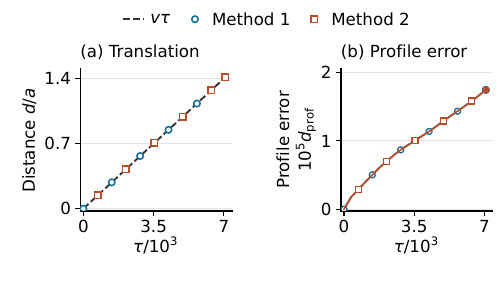}
\caption{A prepared neutral trimer translates with little deformation.
(a) Displacement in lattice units, $d/a$, follows $v\tau$, with dimensionless speed $v=2\times10^{-4}$ and time $\tau=E_0t/\hbar$; the physical speed is $vE_0a/\hbar$.
(b) The global profile error $d_{\rm prof}$ compares the evolved state with a translated initial profile; filled endpoints use an exact lattice shift by $(1,-1)a$.
The two methods are specified in the Supplemental Material.
The total travel, $1.4a$, is about $9\%$ of the $16a$ distance between the outer $A$ lobes.}
\label{fig:prepared-motion}
\end{figure}

The same cancellation brings a stability question: a neutral flag molecule can unwind through local states outside the flag.
We therefore allow all local amplitudes to vary in a mean field calculation, retaining a separate wave function on each site.
Both pairs and trimers survive as numerical local minima in both multiplets.
The representation selects the neutral member among these available molecules, as Figure~\ref{fig:selection} shows.
Figures~\ref{fig:binding} and \ref{fig:prepared-motion} then establish its energetic survival and short distance motion.

\sec{Why the neutral composition changes}
The representation labels $(p,q)$ determine the two gyroscopic weights.
A flag state can be written as $\psi_{p,q}=v^{\otimes p}\otimes\bar w^{\otimes q}$: $p$ copies of $v$ and $q$ copies of the complex conjugate of $w$.
The phase accumulated as the state changes, its Berry phase, adds over the copies.
Conjugation reverses its sign, so the two winding contributions are weighted by $p$ and $-q$.
These weights determine the transverse force per unit velocity.

Let $C_v,C_w$ be the integer windings of the two directions.
We label the elementary textures $(C_v,C_w)=(-1,0)$ and $(0,-1)$ by $A$ and $B$.
A molecule containing $n_A$ and $n_B$ such windings has normalized gyroscopic charge
\begin{equation}
 C=pC_v-qC_w=-pn_A+qn_B \,.
 \label{eq:selection}
\end{equation}
The continuum gyroscopic coefficient is $2\pi\hbar n_0C$ in the Berry convention of the Supplemental Material, with $n_0=a^{-2}$ the site density and $a$ the lattice spacing.
Neutrality means $C=0$, or $pn_A=qn_B$.
Equal weights in $(1,1)$ compensate in an $AB$ pair.
Doubling the weight of $B$ in $(1,2)$ requires two $A$ constituents, giving an $A_2B$ trimer.
Changing an ordinary spin length rescales a single Berry weight; changing the relative weights here changes the composition rule itself.

The dynamical consequence follows by considering an isolated molecule in a conservative continuum with translation symmetry.
The first spatial moment of its gyroscopic charge density is conserved~\cite{PapanicolaouTomaras,BraunerYamamotoYokokura,SornSchmalianGarst}.
A rigid shift changes that moment by the displacement times the total charge.
A neutral molecule can therefore translate without changing the conserved quantity.
This is the conservative counterpart of cancelling the gyroscopic term that produces Hall deflection under a drive.
Whether a bound molecule actually translates depends on its interactions and preparation; below we test that motion directly.

\sec{Holding the energy fixed}
The copy construction also explains how the energy can be held fixed while the gyroscopic weights change.
We choose interactions that couple to the \emph{average} over each set of copies.
For a flag state, averaging $p$ identical copies of $v$ returns the same local orientation regardless of $p$; the same holds for the $q$ copies associated with $w$.
The exchange, local field, and attraction are all normalized in this way.
Their energies depend on the flag directions, while the Berry phases still add over copies.

Writing $\Psi_{p,q}[v,w]$ for the product of the local flag states over the lattice, the resulting Hamiltonians obey
\begin{equation}
 \langle\Psi_{p,q}[v,w]|\widehat H_{p,q}|\Psi_{p,q}[v,w]\rangle
 =E_{\rm flag}[v,w] \,.
 \label{eq:coherentenergy}
\end{equation}
The right hand side is the same function for every positive $(p,q)$ at fixed couplings.
The equality concerns flag states; allowing additional local amplitudes gives different enlarged energy landscapes in the two multiplets.
The operator construction and the explicit lattice energy are given in the Supplemental Material~\cite{Supplemental}.

The energy contains four ingredients familiar from magnetic texture models.
Exchange with coefficient $\rho$ favors similar order on neighboring sites.
An additional stiffness $\alpha$ penalizes rapid spatial bending and opposes collapse into a sharp core.
A local anisotropy with scale $h$ favors a uniform polarized state.
Finally, an attraction of strength $\eta$ lowers the energy where the two winding species overlap.
For the corresponding periodic Hamiltonians, the polarized state is the exact quantum ground state and the gap is $h$ throughout the parameter range used here~\cite{Supplemental}.
The textures are therefore studied above a well defined gapped vacuum.

\sec{Binding without topological protection}
A flag describes the most ordered local states, but a general local wave function has eight or fifteen complex amplitudes.
After fixing its normalization and removing its arbitrary overall phase, these amplitudes form a projective space.
A smooth two dimensional texture in that space has a single integer degree.
For a flag texture this degree is
\begin{equation}
 C_{\rm full}=pC_v-qC_w=C \,.
 \label{eq:ambient-degree}
\end{equation}
The identity follows by evaluating the Berry curvature of the full wave function on the flag states~\cite{KimOh,AffleckBykovWamer,EmbedHatcher}.
Consequently, cancelling the gyroscopic force also removes topological protection: the neutral molecule can be deformed continuously to the vacuum while its distant boundary stays fixed.
The local wave function can leave the flag during this deformation and remain well defined.

The exchange and stiffness terms give an energy cost to these departures from flag order.
We test whether this cost, together with the attraction, sustains a molecule when every local amplitude is free to relax.
Both compositions converge to stationary configurations in both representations.
Independent searches for their softest distortions give positive energy curvatures after the two exact global internal phase rotations are removed~\cite{Supplemental}.
These calculations support numerical local minima of the mean field energy.
Figure~\ref{fig:binding}(a,b) shows the neutral trimer after this relaxation: the two outer $A$ lobes and the central $B$ lobe remain together.

For binding, we compare the molecule with separately optimized fragments.
The closest tested breakup of the neutral trimer is an $A_2$ dimer and a single $B$.
At $\eta/h=4$, the binding is about $2.1\times10^{-3}E_0$, or $0.17\%$ of the trimer's excitation energy and $2.6$ times the gap $h$.
Here, $E_0$ is the common energy unit.
This modest binding is well resolved: changing the lattice size shifts the trimer energy by less than $4.4\times10^{-8}E_0$~\cite{Supplemental}.
The other three composition and representation combinations also lie below their tested breakup comparisons.

Figure~\ref{fig:binding}(c,d) follows the neutral trimer through ten attraction strengths, $3.99\le\eta/h\le4.20$, a range of about $5\%$.
Its binding grows while its easiest shape deformation becomes softer.
The plotted shape stiffness is the second derivative of the energy along that deformation; translations and rotation are checked separately and remain in the stability analysis.
Positive binding and computed low curvatures persist across this sampled range and under independent changes of both stiffnesses, including $\pm10\%$ in $\alpha$~\cite{Supplemental}.
The molecule thus survives more than a single parameter choice, with a shape that becomes increasingly flexible toward the upper end of the scan.

\sec{Seeing the cancellation in motion}
Figure~\ref{fig:selection}(b,c) compares one stretched flag pair released into each multiplet.
The initial patterns and their energies agree, so the relative constituent response probes the changed gyroscopic weights.
The two species move together in $(1,1)$, whereas $A$ travels about twice as far as $B$ in $(1,2)$.
For flag states in the continuum, conservation of the first moment gives $p\,\Delta\bm R_A=q\,\Delta\bm R_B$ for this pair, and hence the displacement ratio $q{:}p$.
The centers measure collective winding distributions and can move through fractions of a lattice spacing.
Their travel is about $0.05a$, compared with a separation of $14a$.
Without the imposed stretch, the midpoint drifts by only $1$--$2\%$ of its perturbed displacement; subtracting this background leaves species displacement ratios of $1.000$ and $1.998$~\cite{Supplemental}.

The trimer supplies the reciprocal comparison in Figure~\ref{fig:selection}(d,e).
Its $A$ center includes both lobes of the winding distribution in Figure~\ref{fig:binding}(a).
The continuum flag prediction is now $2p\,\Delta\bm R_A=q\,\Delta\bm R_B$: the species move at a $1{:}2$ ratio in $(1,1)$ and together in $(1,2)$.
In all four panels, allowing every local amplitude to evolve closely follows the calculation restricted to flag states.
The pair differences are at the percent level and the trimer differences smaller~\cite{Supplemental}.
The pair trajectories slow in both descriptions during this early response; the mechanism of that slowing remains open.

For a direct test of translation, we separately prepare a profile whose evolution is approximately a uniform spatial shift at a prescribed velocity, then release it without a drive.
Figure~\ref{fig:prepared-motion} shows the neutral trimer evolving with all local amplitudes free.
Two independent numerical methods agree on its displacement and deformation.
The dimensionless profile error $d_{\rm prof}$ compares the evolved texture with a shifted copy of the initial one; it combines local differences over the whole lattice, with arbitrary wave function phases removed~\cite{Supplemental}.
After moving $(1,-1)a$, it is about $1.75\times10^{-5}$.
For comparison, a texture that remained unmoved would give $d_{\rm prof}=5.26$ against the same translated target.
This travel of $\sqrt2a\simeq1.4a$ is about $9\%$ of the $16a$ separation between the outer lobes.
The neutral pair also admits a prepared profile that translates with little deformation~\cite{Supplemental}.
These are demonstrations over short distances; propagation over many molecule sizes and the lifetime against quantum unwinding remain open.

The local quantum multiplet provides a way to change the neutral composition of a skyrmion molecule while preserving the entire energy landscape of its flag order.
The same geometry also exposes the neutral molecule to unwinding, making energetic survival part of the selection problem.
The pair and trimer establish this mechanism within one controlled family of models.
Local representations built from orbital occupations offer a connection to multicomponent magnets and atomic systems~\cite{GorshkovTwoOrbital,ScazzaOrbitalExchange}; the construction and its experimental scope are discussed in the Supplemental Material.
The next physical question is which accessible interactions can support this composition control while retaining the molecules' binding and motion.

\sec{Acknowledgments}
The author thanks the NSF Institute for Artificial Intelligence and Fundamental Interactions (IAIFI) and the Department of Physics at Northeastern University for hospitality during his sabbatical, when much of this research was undertaken. He also thanks Izak Snyman for comments on an early draft.
VJ is supported by the South African Research Chairs Initiative of the Department of Science, Technology, and Innovation and the National Research Foundation (grant 78554).
GPT-5.6, GPT-6, and Claude Fable 5.1 assisted in this work.

\sec{Data availability}
The reference fields, operators, numerical records, and verification code are supplied in the companion archive \texttt{numerical\_data.zip}, together with the figure data and plotting scripts, at \href{https://doi.org/10.5281/zenodo.22959390}{doi:10.5281/zenodo.22959390}.

\end{document}


\title{Supplemental Material for \texorpdfstring{\\}{ }``Quantum Representation Selects Neutral Skyrmion Molecules''}
\author{Vishnu Jejjala}
\noaffiliation
\date{September 2026}
\maketitle

\section*{How to read this supplement}
The Letter asks how the local quantum multiplet selects the composition of a molecule with zero gyroscopic charge, and whether that molecule can remain bound and move.
Section~\ref{sec:geometry} derives the cancellation rule and explains why the same rule removes topological protection in the full local state space.
Sections~\ref{sec:family} and \ref{sec:vacuum} construct a family with exactly the same energy for every flag texture and prove that its polarized vacuum is the quantum ground state with a finite gap.
Section~\ref{sec:methods} defines the numerical tests; Sections~\ref{sec:statics} and \ref{sec:coherent-bridge} apply them to binding, parameter changes, and relaxation away from flag order.
Section~\ref{sec:matched-motion} compares the response to identical deformations in the two representations.
Sections~\ref{sec:cp7-travel}--\ref{sec:charged-corrector} test separately prepared translating profiles and the translation reaction of their charged partners.
Section~\ref{sec:encoding} explains a local occupation encoding, and Section~\ref{sec:data} collects the evidence and the limits of the conclusions.

\section*{Terminology, notation, and units}
A \emph{flag state} is an $SU(3)$ coherent state: the simplest classical order available in the representation, specified by two orthogonal complex line directions $v,w$.
The local flag state is $\psi_{p,q}$; the Letter denotes its product over lattice sites by $\Psi_{p,q}[v,w]$.
A \emph{full state} allows all eight or fifteen local complex amplitudes to vary, with unit norm and an irrelevant overall phase.
The ansatz for the lattice remains a product of these local states, which is a mean field treatment; allowing full local states does not introduce entanglement between sites.
Throughout, \emph{neutral} means zero gyroscopic charge.
The term \emph{coherent embedding} retains its conventional mathematical meaning for the map from a flag to its local quantum ray.

\begin{description}
\item[Representations and winding species]
The eight state multiplet is $(p,q)=(1,1)$, with full ray space $\mathbb{CP}^{7}$; the fifteen state multiplet is $(1,2)$, with full ray space $\mathbb{CP}^{14}$.
The texture species $A$ and $B$ have line windings $(C_v,C_w)=(-1,0)$ and $(0,-1)$.
The root convention used in some tables is $\bm k=(C_v,-C_w)$.
In fragment comparisons, $D=A_2$ denotes a dimer and $M=AB$ a mixed pair; the trimer is $A_2B$.
\item[Operators and energy]
$F^a,B^a$ are averaged local generators, with generator index $a$; $B^a$ is an operator, whereas plain $B$ names a texture species. Likewise, $D^a$ in Equation~\eqref{eq:FB} is an operator, distinct from the dimer $D$.
The diagonal matrices $\mathsf A,\mathsf C$ define the attraction and do not label textures.
$E_{\rm flag}$ denotes the common flag energy; it is the quantity called $E_{\rm coh}$ in numerical records.
The positive coefficients $w_F,w_B$ set the strengths of the two constituent exchange terms and are unrelated to the vector $w$.
\item[Spatial and stability observables]
$N$ is the number of sites along each side of the square lattice.
The centers $\bm R_A,\bm R_B$ in the Letter are the moment spectral centers $\bm R_v,\bm R_w$ defined in Section~\ref{sec:matched-motion}; each includes all lobes of its species.
A Hessian curvature measures the quadratic energy cost of a small deformation.
$\lambda_{\rm shape}$ is the lowest computed internal shape curvature after identifying translations and orientation.
\end{description}

Lengths are in lattice spacings $a$, energies in an overall Hamiltonian scale $E_0$, and time in $\hbar/E_0$.
The tabulated couplings are dimensionless: the physical Hamiltonian is $E_0\widehat H$ evaluated at those couplings, with no material calibration of $E_0$ assumed.
The dimensionless time is $\tau$; $a=E_0=1$ in the numerical records.
The continuum Berry action has site density $n_0=a^{-2}$. Berry forms and gyroscopic coefficients below use $\hbar=1$; restoring units multiplies the coefficient $G$ by $\hbar$, as in the Letter.
The lattice spacing is fixed: increasing $N$ tests the box size, rather than a continuum limit.
The reference parameters are
\begin{equation}
 \rho=0.2 \,,\quad\alpha=4 \,,\quad h=0.0008 \,,\quad
 w_F\simeq0.0431765 \,,\quad w_B\simeq0.0730760 \,,
 \label{eq:parameters}
\end{equation}
with $\eta/h=4$ unless varied.
Here $\rho$ penalizes gradients, $\alpha$ opposes collapse through a higher derivative cost, and $h$ sets the local field scale and the exact vacuum gap of the two studied members over the range proved in Section~\ref{sec:vacuum}.
The weights above are rounded; the numerical evidence retains the floating point values actually used.
Displayed results are rounded to the precision needed for each comparison; the numerical archive retains the stored precision. Energy differences, ratios, and convergence diagnostics are evaluated before rounding. Upper and lower bounds are rounded outward, and small departures from unity are stated explicitly where needed.

\section{Representation weights, topology, and translation}
\label{sec:geometry}
Two questions share one mathematical answer: which winding combinations have zero gyroscopic charge, and which can unwind when all local amplitudes are allowed?
The representation assigns an integer weight to each independent winding. Their weighted sum determines both quantities.
We first state this result for a regular coherent orbit and then give the $SU(3)$ conventions used in the calculations.
\subsection{Regular orbits and the coherent state embedding}
Let $G$ be compact, simply connected, and semisimple of rank $r$, with maximal torus $T$.
A regular integral highest weight $\lambda$ has strictly positive Dynkin labels
$\ell_i=\langle\lambda,\alpha_i^\vee\rangle$.
The stabilizer of its highest weight ray is $T$, so its flag orbit is $G/T$.
The fibration $T\to G\to G/T$ gives $\pi_1(G/T)=0$ and $\pi_2(G/T)\simeq\mathbb Z^r$.
Choose the simple root spheres as generators of this integer charge lattice.
For a local normalized state $\psi$, the action one-form is $\Theta=\ii\langle\psi|\dd\psi\rangle$; define the real Berry connection $\mathcal A=-\Theta$ and positive curvature $\Omega=\dd\mathcal A=-\dd\Theta$.
Its periods are $2\pi \ell_i$, the standard highest weight symplectic geometry~\cite{KimOh,AffleckReview}.
For a smooth texture approaching a fixed ray at spatial infinity,
\begin{equation}
 G_\lambda(\bm k)=n_0\int_{S^2}\phi^*\Omega
 =2\pi n_0\sum_{i=1}^r \ell_i k_i \,,
 \qquad
 \mathcal N_\lambda=\ker_{\mathbb Z}(\bm\ell)
 =\{\bm k\in\mathbb Z^r:\bm\ell\cdot\bm k=0\} \,.
 \label{eq:general}
\end{equation}
The image of the integer pairing is $\gcd(\ell_1,\ldots,\ell_r)\mathbb Z$, and its kernel has rank $r-1$.
A common rescaling of all Dynkin labels leaves it unchanged, while a change in their ratios generally changes it.
For $SU(N)$ the neutral lattice has rank $N-2$; for example $(1,1,-1)$ is charged for $SU(4)$ labels $(1,1,1)$ and neutral for $(1,1,2)$.
This example is a kinematic consequence of Equation~\eqref{eq:general}, not an additional computed texture.

Let $V_\lambda$ be the representation space and
$\iota_\lambda:G/T\hookrightarrow\mathbb{CP}^{\dim V_\lambda-1}$ its highest weight coherent embedding.
The pullback of the projective hyperplane class $c_1(\mathcal O(1))$ has periods $\ell_i$.
Both target spaces are simply connected, so their natural Hurewicz maps identify $\pi_2$ with $H_2$~\cite{Hatcher}.
The induced homomorphism is therefore
\begin{equation}
 (\iota_\lambda)_*:\mathbb Z^r\longrightarrow\mathbb Z \,,
 \qquad \bm k\longmapsto\bm\ell\cdot\bm k \,.
 \label{eq:generalembedding}
\end{equation}
Its kernel is precisely $\mathcal N_\lambda$.
Every neutral coherent class is nullhomotopic in the full local projective space with the asymptotic ray held fixed.
This statement concerns the configuration space already at product state level; it neither requires intersite entanglement nor describes an energetic escape path. On a finite lattice, topology may also change at its cutoff; the homotopy statement refers to smooth texture maps.

\subsection{Explicit flag conventions}
For $SU(3)$ write $\lambda=(p,q)$ with $p,q>0$ and
\begin{equation}
 d_{p,q}=\frac{(p+1)(q+1)(p+q+2)}2 \,,
 \qquad
 M=V\Lambda_{p,q}V^\dagger \,,
 \qquad
 \Lambda_{p,q}=\frac13\operatorname{diag}(2p+q,q-p,-p-2q) \,.
 \label{eq:flagmoment}
\end{equation}
Here, $V\in SU(3)$; its first and third columns are denoted $v,w$.
They satisfy $v^\dagger w=0$.
The normalized coherent lift
$\psi_{p,q}=v^{\otimes p}\otimes\bar w^{\otimes q}$ belongs to the Cartan product summand of
$\operatorname{Sym}^p\bm3\otimes\operatorname{Sym}^q\overline{\bm3}$, and
\begin{equation}
 \langle\psi_{p,q}|\dd\psi_{p,q}\rangle=pv^\dagger\dd v-qw^\dagger\dd w \,,
 \qquad
 C_z=\frac1{2\pi}\int[-\ii\,\dd(z^\dagger\dd z)] \,,
 \qquad (k_1,k_2)=(C_v,-C_w) \,.
 \label{eq:topology}
\end{equation}
The middle eigenline has charge $C_2=-C_v-C_w$, hence the complete eigenline vector is
$(C_1,C_2,C_3)=(k_1,k_2-k_1,-k_2)$.
In this orientation
\begin{equation}
 C_{\rm full}=pC_v-qC_w=pk_1+qk_2 \,,
 \qquad
 (k_1,k_2)_{\rm primitive}=\left(\frac qg,-\frac pg\right) \,,
 \qquad g=\gcd(p,q) \,.
 \label{eq:primitive}
\end{equation}
A root-$12$ sphere mixes only the first two eigenlines and carries $k_2=0$; a root-$23$ sphere mixes only the last two and carries $k_1=0$.
Their curvature periods are respectively $2\pi p$ and $2\pi q$.
The equality for $C_{\rm full}$ is exact for flag fields.
For arbitrary rays in $\mathbb{CP}^{d-1}$, principal constituent projectors may still be defined when their largest eigenvalues are isolated, but their local Berry densities do not give an exact decomposition of the full projective density.
The latter must then be evaluated independently.

\subsection{Conserved moment and its scope}
For localized smooth planar fields, let $q_\Omega(\bm x)\dd x\wedge\dd y=n_0\phi^*\Omega$ and define
\begin{equation}
 G=\int q_\Omega\,\dd^2x \,,
 \qquad \bm D=\int\bm xq_\Omega\,\dd^2x \,,
 \qquad \bm P=(D_y,-D_x) \,.
 \label{eq:DP}
\end{equation}
The integrals require sufficient decay and an asymptotic state independent of position.
Translation symmetry of the Hamiltonian conserves $\bm D$; this moment law for symplectic targets and its magnetic applications are established results~\cite{Papanicolaou,Brauner,Lee}.
An active translation by $\bm a$ changes $\bm D$ by $G\bm a$.
Thus, a nonzero $G$ excludes nonzero rigid translation in a conservative, continuously translation invariant continuum, whereas $G=0$ removes that obstruction.
Internal motion of a charged texture is not excluded.
Nor does nonzero gyrocharge imply spatial pinning: an external force can produce transverse motion, and a boundary or spatially varying field removes the translation symmetry used in the conserved moment argument.
Conversely, zero gyrocharge is not sufficient for binding, lattice depinning, or asymptotically stable transport.
For an ordinary $SU(2)$ spin, varying the spin length rescales the one Berry weight and leaves the cancellation rule unchanged; neutral pairs in the zero winding class can still exist.
A regular higher rank flag instead admits a nonzero winding vector in the kernel of the weighted pairing.
This differs from $\mathbb{CP}^{N-1}$, whose second homotopy group has only one integer generator even when $N>2$.

For a flag, let $b_z$ be the signed density defined by
$b_z\,\dd x\wedge\dd y=-\ii\,\dd(z^\dagger\dd z)/(2\pi)$.
The root densities are $\rho_1=b_v$ and $\rho_2=-b_w$, so their integrals are $k_1=C_v$ and $k_2=-C_w$.
Define complete signed centers
$\bm R_j=k_j^{-1}\int\bm x\rho_j\dd^2x$ for $j=1,2$.
Each center includes all lobes of its species.
For the positive primitive orientation in Equation~\eqref{eq:primitive},
\begin{equation}
 G_*=2\pi n_0\frac{pq}{g} \,,
 \qquad \bm s=\bm R_2-\bm R_1 \,,
 \qquad \bm D=-G_*\bm s \,,
 \qquad \bm P=G_*\widehat{\bm z}\times\bm s \,.
 \label{eq:coherentmomentum}
\end{equation}
For the neutral trimer's opposite charge orientation $(-2,+1)$ at $(p,q)=(1,2)$, the flag limit gives
$\bm P=-4\pi n_0\widehat{\bm z}\times(\bm R_w-\bm R_v)$.
The exact continuum definition outside the flag remains Equation~\eqref{eq:DP} with the full projective density.
Small constituent purity errors alone do not bound errors in its spatial derivatives or first moment.
Neither center formula is an exact continuously generated momentum on the physical square lattice.
The full space response and releases below use the local projective symplectic form directly.

\section{One normalized constituent Hamiltonian family}
\label{sec:family}\label{sec:model}
To attribute a changed response to Berry weights, the energy of the same flag texture must remain fixed when the representation changes.
A $(p,q)$ flag contains $p$ copies of $v$ and $q$ conjugate copies of $w$.
Berry phases add over these copies, whereas the Hamiltonian below couples to their averages.
This keeps the flag energy fixed while allowing the full local state spaces and their energies away from the flag to differ.
\subsection{Local representation and operators}
Let $U_{p,q}:V_{p,q}\to(\mathbb C^3)^{\otimes p}\otimes
(\overline{\mathbb C^3})^{\otimes q}$ be an isometry onto the Cartan product summand.
Its tensors are symmetric within each set of indices and annihilated by every fundamental--antifundamental trace contraction.
The vacuum is the highest weight ray represented by
$|1\rangle^{\otimes p}\otimes|\bar3\rangle^{\otimes q}$.
Write $f_r^a=t^a$ on fundamental factor $r$ and $b_s^a=-(t^a)^*$ on antifundamental factor $s$, with
$\operatorname{tr}(t^at^b)=\delta^{ab}/2$.
The compressed averages are
\begin{equation}
 F^a=U_{p,q}^\dagger\left(\frac1p\sum_{r=1}^p f_r^a\right)U_{p,q} \,,\qquad
 B^a=U_{p,q}^\dagger\left(\frac1q\sum_{s=1}^q b_s^a\right)U_{p,q} \,,\qquad
 T^a=pF^a+qB^a \,.
 \label{eq:constituents}
\end{equation}
Permutation symmetry makes either average equal to the compression of any one factor of that type.
The averages $F,B$ transform together as parts of one irreducible local moment. They are not two independent spins that can fluctuate separately on a site.

Put $\mathsf A=\operatorname{diag}(0,1,1)$ and $q_F=\operatorname{diag}(0,1,3)$ on a fundamental factor,
and $\mathsf C=\operatorname{diag}(1,1,0)$ and $q_B=\operatorname{diag}(3,2,0)$ on an antifundamental factor.
These matrices measure departures from the vacuum directions. Bars denote averages over factors of the corresponding type. Define
\begin{equation}
 \begin{split}
 K&=U_{p,q}^\dagger\overline{\mathsf A}\overline{\mathsf C} U_{p,q} \,,\\
 \bar Q&=U_{p,q}^\dagger\left(\frac1p\sum_r q_{F,r}+\frac2q\sum_s q_{B,s}\right)U_{p,q}
 =\frac{14}{3}I-(F^3+2B^3)-\frac5{\sqrt3}(F^8+2B^8) \,.
 \end{split}
 \label{eq:Kdefinition}
\end{equation}
The product in $K$ is taken before compression; it is generally not the product of two compressed operators.
Both onsite operators preserve the global Cartan symmetries, $0\le K\le I$, and $K|0\rangle=\bar Q|0\rangle=0$.
For general $(p,q)$, $\bar Q$ is a constituent field operator rather than a linear total Cartan field.
This normalization is part of the definition of the family.
At $(1,2)$ it is exactly $Q=14I/3-T^3-5T^8/\sqrt3$.
The relative factor $2$ in the antifundamental field average is a chosen anisotropy, held fixed as $q$ changes. It supplies the coefficients of $\Phi$ below; it is not a representation dependent rescaling of the total generator.

The vacuum subtracted quantum Hamiltonian is
\begin{equation}
 \widehat H^{(p,q)}_\eta=
 \sum_{(ij)}c_{ij}\left[\frac{w_F+w_B}{3}
 -w_F F_i^aF_j^a-w_B B_i^aB_j^a\right]
 +\sum_i(h\bar Q_i-\eta K_i) \,.
 \label{eq:latticeH}
\end{equation}
Bonds are unordered. Nearest neighbor, axial distance two, and diagonal coefficients are
$2\rho+16\alpha$, $-2\alpha$, and $-4\alpha$, respectively.
Repeated generator indices are summed.
The exact theorem below assumes $w_F,w_B,\rho,\alpha\ge0$, $h>0$, and a periodic square lattice of side at least five.
A vacuum collar is imposed only in specified product state calculations.

For direct construction of $(1,1)$, identify $\bm3\otimes\bar{\bm3}$ with complex $3\times3$ matrices and take the eight columns of $U_{1,1}$ to be $\sqrt2\operatorname{vec}(t^a)$, in a consistent vectorization convention.
For $(1,2)$ one may compress the $27$-dimensional tensor product or use
\begin{equation}
 D^a=d^{abc}T^bT^c \,,\qquad
 F^a=\frac{11}{24}T^a+\frac14D^a \,,\qquad
 B^a=\frac{13}{48}T^a-\frac18D^a \,.
 \label{eq:FB}
\end{equation}
There are two independent sets of operators in $\operatorname{End}(V_{1,2})$ that transform as the adjoint; $T,D$ span them, and
the vacuum matrix elements $\langle D^3\rangle=13/12$ and $\langle D^8\rangle=-31/(12\sqrt3)$ fix the coefficients.
The numerical evidence specifies the basis of the fifteen state operators explicitly.

\subsection{Exact product energy and common flag restriction}
The generator averages determine two reduced density matrices.
On a flag each is a rank one projector; away from the flag their loss of purity gives an additional local energy cost.
Writing the full product energy in these variables makes that cost and the common flag restriction explicit.
For normalized rays put $m_{X,i}^a=\langle\psi_i|X^a|\psi_i\rangle$, $X=F,B$, and
\begin{equation}
 R_{v,i}=\frac I3+2m_{F,i}^at^a \,,\qquad
 R_{w,i}=\frac I3-2m_{B,i}^at^a \,.
 \label{eq:marginals}
\end{equation}
These are the fundamental one factor density matrix and the complex conjugate of an antifundamental one factor density matrix.
The conjugation fixes the sign of $C_w$.
Both are positive with unit trace, hence $|\bm m_X|^2\le1/3$.
With $Lf_i=4f_i-\sum_{j\sim i}f_j$ and $s_0=8\rho+40\alpha$, the exact product energy is
\begin{equation}
 \begin{split}
 E[\psi]={}&\sum_{X=F,B}w_X\left[
 \frac{s_0}{2}\sum_i\left(\frac13-|\bm m_{X,i}|^2\right)
 +\rho\sum_{i,\mu=x,y}|\bm m_{X,i+\hat\mu}-\bm m_{X,i}|^2
 +\alpha\sum_i|L\bm m_{X,i}|^2\right]\\
 &+\sum_i\langle\psi_i|h\bar Q-\eta K|\psi_i\rangle \,.
 \end{split}
 \label{eq:product}
\end{equation}
The first term measures the reduction in constituent purity and is present whenever a full state leaves the flag.
Joint saturation of the two bounds forces every single factor reduction to be pure, so the ambient tensor factors as
$v^{\otimes p}\otimes\bar w^{\otimes q}$; trace cancellation requires $v^\dagger w=0$.
When both weights and $s_0$ are positive, the local purity cost vanishes precisely on the flag.
No nonlinear replacement of $\langle K\rangle$ is made in Equation~\eqref{eq:product}.

For a flag state, the moments depend only on $P_v=vv^\dagger$ and $P_w=ww^\dagger$ and are independent of $(p,q)$.
Using $|\Delta\bm m|^2=\|\Delta P\|_F^2/2$ gives
\begin{equation}
 \begin{split}
 E_{\rm flag}[v,w]={}&\sum_{X=v,w}\frac{w_X}{2}
 \left[\rho\sum_{i,\mu}\|P_{X,i+\hat\mu}-P_{X,i}\|_F^2
 +\alpha\sum_i\|LP_{X,i}\|_F^2\right]\\
 &+\sum_i\left[h\Phi_i-\eta(1-P_{v,i}^{11})(1-P_{w,i}^{33})\right] \,,\\
 \Phi={}&P_v^{22}+3P_v^{33}+6P_w^{11}+4P_w^{22} \,,
 \qquad(w_v,w_w)=(w_F,w_B) \,.
 \end{split}
 \label{eq:coherent}
\end{equation}
This is exactly representation independent at fixed couplings, whereas Equation~\eqref{eq:topology} has Berry weights $(p,q)$.
Changing representation also changes the normal directions and the off orbit energy; unrestricted energies and trajectories are not asserted to coincide.

\subsection{Physical roles of the interaction terms}
The physical requirements are distinct: a field selects the vacuum, a higher derivative stiffness prevents shrinking, and the onsite term favors overlap of the two winding species.
For a localized flag continuum trial shape dilated by a dimensionless factor $R$, Equation~\eqref{eq:coherent} has the schematic scaling
\begin{equation}
 E(R)=E_2+\frac{E_4}{R^2}+R^2 E_{\rm loc} \,,\qquad
 E_{\rm loc}=\int\dd^2x\,[h\Phi-\eta(1-P_v^{11})(1-P_w^{33})] \,.
 \label{eq:scaling}
\end{equation}
Here, $E_2$, $E_4$, and $E_{\rm loc}$ refer to the undilated continuum profile.
For $E_4,E_{\rm loc}>0$, their competition selects a finite scale; this scaling argument is not an exact lattice dilation or a binding theorem.
In the tested family the onsite attraction supplies the overlap gain, and the separately relaxed candidate at zero attraction does not bind against the explicit comparison in Section~\ref{sec:statics}.
Table~\ref{tab:representations} summarizes the two representations and their neutral compositions.

\begin{table}[ht]
\caption{The common family and its neutral composition, in the orientation used numerically.}
\label{tab:representations}
\begin{ruledtabular}
\begin{tabular}{cccc}
$(p,q)$ & Local ray space & Primitive neutral $\bm k$ & $C_{\rm full}$ of $\bm k=(-1,+1)$\\
$(1,1)$ & $\mathbb{CP}^{7}$ & $(-1,+1)$ & $0$\\
$(1,2)$ & $\mathbb{CP}^{14}$ & $(-2,+1)$ & $+1$\\
\end{tabular}
\end{ruledtabular}
\end{table}

\section{Positive quantum exchange and exact vacuum bounds}
\label{sec:vacuum}
The vacuum proof establishes the ground state above which the textures are excitations.
We prove that the polarized product vacuum is the exact quantum ground state and has a finite excitation gap.
The proof first expresses the exchange as positive operators, then bounds the onsite field and attraction.
It applies to the periodic lattices specified in Section~\ref{sec:family}; the existence and binding of textures are separate numerical questions.
\subsection{Exchange positivity before compression}
For one fundamental or antifundamental factor, let $\bm s_i$ be its generators and $\mathsf S_{ij}$ exchange identical factors on adjacent sites.
Give each fundamental copy weight $w_F/p$ and each antifundamental copy weight $w_B/q$, collectively denoted $\gamma_o$.
The uncompressed exchange is
\begin{equation}
 \widetilde H_{\rm ex}=\sum_o\gamma_o\left\{
 \rho\sum_{\langle ij\rangle}(I-\mathsf S_{ij}^{(o)})
 +\alpha\sum_i\left[\left(\sum_{j\sim i}\bm s_j^{(o)}-4\bm s_i^{(o)}\right)^2-20I\right]\right\} \,.
 \label{eq:stars}
\end{equation}
The permutation terms are positive.
For a star, combine the four neighbors into the irreducible representation $\mu$ and then add the center to obtain $\nu$.
The square has eigenvalue $5C_2(\mu)+80/3-4C_2(\nu)$, where
$C_2(p,q)=(p^2+q^2+pq+3p+3q)/3$.
For four fundamentals the allowed $\mu$ are $(4,0),(2,1),(0,2),(1,0)$; the minimum square eigenvalues are $20,20,22,20$.
Conjugation gives the antifundamental result. Each subtracted star is positive.
Expansion gives
\begin{equation}
 \sum_i\left(\sum_{j\sim i}\bm s_j-4\bm s_i\right)^2
 =\frac{80N_s}{3}I-16\sum_{\rm NN}\bm s_i\cdot\bm s_j
 +2\sum_{\rm axial\,2}\bm s_i\cdot\bm s_j
 +4\sum_{\rm diagonal}\bm s_i\cdot\bm s_j \,.
 \label{eq:expansion}
\end{equation}
Compression by $\bigotimes_iU_{p,q}$ gives the exchange in Equation~\eqref{eq:latticeH}, including its scalar subtraction.
Squaring already compressed constituent operators would not prove this identity.
The star argument uses fundamental and antifundamental $SU(3)$ factors; it is not a theorem for arbitrary compact groups.

\subsection{Bounds for every regular \texorpdfstring{$(p,q)$}{(p,q)}}
On an ambient number basis write $a=\overline{\mathsf A}\in\{0,1/p,\ldots,1\}$ and $c=\overline{\mathsf C}\in\{0,1/q,\ldots,1\}$.
Since $q_F\ge\mathsf A$ and $q_B\ge2\mathsf C$,
\begin{equation}
 \bar Q_{\rm amb}-4\overline{\mathsf A}\overline{\mathsf C}\ge a+4c-4ac=a+4c(1-a) \,.
 \label{eq:ambient-bound}
\end{equation}
Except at the unique vacuum $a=c=0$, this is at least $\delta_{p,q}=\min(1/p,4/q)$.
Compression and positive exchange imply
\begin{equation}
 \widehat H^{(p,q)}_\eta\ge h\delta_{p,q}\sum_i(I-|0\rangle_i\langle0|) \,,
 \qquad 0\le\eta\le4h \,.
 \label{eq:gapbound}
\end{equation}
The global first root and second root descendants have zero exchange, $K=0$, and energies $h/p$ and $4h/q$.
They attain the bound, so the finite volume gap is exactly $h\delta_{p,q}$ throughout this interval.
It is $h$ for both studied representations, but is not uniformly $h$ over all $(p,q)$.

A stronger compressed inequality is
\begin{equation}
 \bar Q\ge7K \,.
 \label{eq:seven}
\end{equation}
First restrict one fundamental--antifundamental pair to its traceless matrix adjoint subspace.
On off diagonal matrix units $q_F+2q_B-7\mathsf A\mathsf C$ has eigenvalues $4,0,0,1,2,0$.
On diagonal amplitudes $a,b,c$ with $a+b+c=0$, its quadratic form is
$6|a|^2-2|b|^2+3|c|^2=|2a-c|^2$.
Every fundamental--antifundamental pair in the Cartan product tensor has zero singlet component, by the trace constraints.
Averaging the pair inequality over all $pq$ pairs proves Equation~\eqref{eq:seven}.
Consequently
\begin{equation}
 \widehat H^{(p,q)}_\eta\ge
 h\left(1-\frac{\eta}{7h}\right)\delta_{p,q}
 \sum_i(I-|0\rangle_i\langle0|) \,,\qquad 0\le\eta<7h \,.
 \label{eq:seven-gap}
\end{equation}
This lower gap bound suffices for a unique polarized vacuum.
The uniqueness endpoint is sharp: at $\eta=7h$ the uniform flags
$(v,w)=(e_2,e_1)$ and $(e_3,e_2)$ also have zero energy, since $\bar Q=7$, $K=1$, and their exchange vanishes.
At larger attraction they lie below the polarized state.

\subsection{A common exact gap interval for the two calculated members}
For $(1,2)$, $Q=\bar Q$ commutes with $K$.
Their joint spectrum is listed in Table~\ref{tab:onsite}.
An exact rational construction uses the symmetrizer $P_s$ on the two antifundamentals and trace map
$C_{a,(i,j,k)}=(\delta_{ij}\delta_{ka}+\delta_{ik}\delta_{ja})/2$.
Then $CC^\dagger=2I_3$ and $P_{15}=P_s-C^\dagger C/2$.
The ambient $Q$ commutes with $P_{15}$; restricting $P_{15}K_{27}P_{15}$ in each $Q$ block gives the table.
The two nontrivial characteristic polynomials are $\kappa^2-7\kappa/8+1/8$ and $\kappa^2-9\kappa/8+1/4$.
\begin{table}[ht]
\caption{Exact joint onsite spectrum for $(1,2)$. Repeated entries retain multiplicity; there are fifteen states.}
\label{tab:onsite}
\begin{ruledtabular}
\begin{tabular}{cc@{\qquad\qquad}cc}
$Q$ & $K$ & $Q$ & $K$\\
0 & 0 & 5 & $(7\pm\sqrt{17})/16$\\
1 & 0 & 6 & $(9\pm\sqrt{17})/16$\\
2 & 0 & 7 & $1,1$\\
3 & $0,3/8$ & 8 & $1$\\
4 & $0,1/2$ & 9 & $1$\\
\end{tabular}
\end{ruledtabular}
\end{table}
\FloatBarrier
The minimum of $(Q-1)/K$ for $K>0$ is $16/3$, reached at $(Q,K)=(3,3/8)$.
Thus, $Q-(\eta/h)K\ge I-|0\rangle\langle0|$ through $\eta/h=16/3$.
For $(1,1)$ the six off diagonal states have $(\bar Q,K)=(0,0),(1,0),(4,0),(7,1),(7,1),(9,1)$.
The two remaining onsite eigenvalues at $x=\eta/h$ are
\begin{equation}
 \lambda_\pm(x)=\frac{14-x\pm\sqrt{x^2-x+7}}3 \,.
 \label{eq:adjoint-onsite}
\end{equation}
The lower one is at least one through $x=38/7$.
Together with the root descendant this establishes the common exact many body gap
\begin{equation}
 \Delta=h \,,\qquad 0\le\eta/h\le16/3 \,,\qquad(p,q)=(1,1),(1,2) \,.
 \label{eq:common-gap}
\end{equation}
The upper endpoint is sufficient, not a proof of the maximal exact gap interval.
These vacuum bounds encompass the texture scan below but do not prove existence, binding, or lifetime of a nonuniform excitation.
\section{Variational and numerical methods}
\label{sec:methods}
Three tests distinguish a numerically stable texture from a field that merely looks stationary: its full energy gradient must vanish, its lowest computed Hessian curvatures must be positive, and its charges and spatial observables must be evaluated independently.
This section fixes the tangent metric and definitions used in those tests, then gives the evolution equation used for releases.
\subsection{Full local rays, gradients, and stationary refinement}
A site carries a normalized $d$-component ray, with horizontal tangent
$\psi_i^\dagger\xi_i=0$ and real inner product $\operatorname{Re}\sum_i\xi_i^\dagger\zeta_i$.
Write $P_i^\perp=I-\psi_i\psi_i^\dagger$ and define the local effective matrix
\begin{equation}
 {\cal A}_i=h\bar Q-\eta K-\sum_{j,X}c_{ij}w_Xm_{X,j}^aX^a \,,\qquad
 G_i={\cal A}_i\psi_i \,,\qquad
 \nabla_iE=2P_i^\perp G_i \,.
 \label{eq:gradient}
\end{equation}
The factor two follows from $\dd E=2\operatorname{Re}\sum_i\delta\psi_i^\dagger G_i$.
The full covariant Hessian gives the change in this gradient under a tangent deformation:
\begin{equation}
 (\mathcal H\xi)_i=
 2P_i^\perp\left[({\cal A}_i-\langle{\cal A}_i\rangle I)\xi_i
 -\sum_{j,X}c_{ij}w_X\,\delta m_{X,j}^a X^a\psi_i\right] \,,
 \qquad
 \delta m_{X,j}^a=2\operatorname{Re}(\xi_j^\dagger X^a\psi_j) \,.
 \label{eq:hessian}
\end{equation}
In particular, the onsite attraction contributes
$-2\eta W_i^\dagger(K-\langle K\rangle_iI)W_i$ in any orthonormal horizontal basis $W_i$.
The formula includes the curvature introduced by keeping each ray normalized.
Finite differences on generic small fields check both formulas to approximately $10^{-10}$ relative error. Separate dense and sparse implementations of their restrictions agree near machine precision.

Collar calculations freeze the outer two site layers to the vacuum and retain all crossing bonds.
Thus, the active dimensions are $2(d-1)(N-4)^2$: $121086$ for the $N=97$ pair in $\mathbb{CP}^7$, and $242172$ in $\mathbb{CP}^{14}$.
Every site is active in periodic calculations.
No final static constraint fixes charges, centers, separation, orientation, reflection, or membership in the flag.
Two exact global Cartan phase directions are removed when independent; their numerical rank is detected before projection.
This quotient does not impose fixed Cartan charges.

Initial fields are formed from embedded root profiles, followed by symmetric polar orthogonalization of their line frame.
Energy relaxation is followed by stationary equation refinement in the full local rays.
The reference trimer uses L-BFGS-B (memory 15, up to 40 line search evaluations) followed by Newton--MINRES refinement.
The stopping quantity reported here is the independently evaluated physical gradient, not the optimizer's termination flag.
On larger boxes and during continuation, nearly free translations make the stationary equations difficult to solve.
We separate those soft translations from local deformations through a Schur complement, then implement the spatial update by a flow retraction.
Translations are restored in accepted updates, the final gradient, and all final physical spectra.
Temporary position conditions are therefore solver coordinates, not constraints defining the reported collar states.

\subsection{Spectral searches and flag Hessians}
An energy minimum must resist deformations in every allowed local direction, including those absent from a flag ansatz.
We therefore evaluate the full Hessian, using a split into flag and normal directions only to accelerate its numerical inversion.
A local generator tangent decomposition supplies six dominant orbit directions and a complementary block of dimension eight for $\mathbb{CP}^7$ or 22 for $\mathbb{CP}^{14}$.
This split constructs a preconditioner; full Hessian actions retain all couplings.
The orbit principal block is treated sparsely, with a positive numerical shift; local inverse blocks precondition the complement.
A floor used in these inverses is a solver parameter, not a lower bound on physical curvature.

Eight low eigenpairs are calculated by solving shifted Hessian equations, typically near $\sigma=-10^{-5}$, with MINRES. Each resulting eigenvector is checked against the complete Hessian action.
Independent LOBPCG searches for the smallest algebraic eigenvalues supply a second test.
The pairs also have independent searches from random starting vectors for their lowest six modes.
The independent $\mathbb{CP}^{14}$ pair search did not converge its seventh and eighth vectors; those are not used, while the separate precise run with eight eigenpairs supplies the reported higher pair values.
For the trimer, independent searches cover the scan endpoints, larger boxes, stiffness controls, and mixed fragment endpoints.
A positive block Gershgorin bound on the normal principal block is also recorded.
Such a bound does not certify its coupled Schur complement, and finite low mode searches are numerical local stability evidence rather than a validated full spectrum inertia theorem.

Flag optimization varies all six flag directions and uses the same Hamiltonian restricted to its coherent embedding.
Its Hessian includes the second fundamental term:
schematically $\mathcal H_{\rm coh}=J_\iota^T\mathcal H J_\iota+
(\nabla E)\cdot\dd^2\iota$.
It is not merely a principal block of the ambient Hessian at a flag stationary point, where the ambient normal force can be nonzero.
Independent automatic differentiation gives relative flag gradient and Hessian action errors $2.0\times10^{-16}$ and $4.3\times10^{-16}$.
Flag trimer curvatures below use the metric induced by its $(1,2)$ lift.

The listed static curvatures are not Hamiltonian mode frequencies.
In the real horizontal coordinates used here, $E=E_*+\tfrac12 u^T\mathcal H u+\cdots$ and linearized TDVP gives
\begin{equation}
 2\hbar\dot u=J_0\mathcal H u \,,\qquad
 J_0=\bigoplus\begin{pmatrix}0&1\\-1&0\end{pmatrix} \,.
 \label{eq:dynamical-generator}
\end{equation}
Here the dot denotes physical time; in dimensionless time $\tau$, the generator is $J_0(\mathcal H/E_0)/2$.
For one canonical pair with curvatures $a,b$, the frequency is $\sqrt{ab}/(2\hbar)$, rather than either curvature alone.
Moreover, the Euclidean phase quotient used for static stability is not automatically a symplectic reduction at fixed charges.
No dynamical spectrum or internal oscillation period is inferred from the tabulated static modes. The release durations and observed slowing are reported directly.

\subsection{Charges, centers, distances, and dynamics}
A molecule's winding, center motion, and profile deformation answer different questions and require distinct observables.
The plaquette phase below measures winding on the lattice; signed centers track each complete winding species, while a distance between local rays tests the whole field.
For any normalized ray field $z$, define the oriented plaquette phase
\begin{equation}
 \phi_\square[z]=\arg\!\left[
 \langle z_1|z_2\rangle\langle z_2|z_3\rangle
 \langle z_3|z_4\rangle\langle z_4|z_1\rangle\right] \,,
 \qquad C[z]=\frac1{2\pi}\sum_\square\phi_\square[z] \,.
 \label{eq:plaquette}
\end{equation}
The phase lies in $(-\pi,\pi]$; nonzero links and a plaquette margin away from the branch cut are checked.
We evaluate the full projective degree independently of the constituent degrees.
For separated principal eigenvalues, raw marginal centers use the principal lines of $R_v,R_w$ in Equation~\eqref{eq:marginals}.
Moment spectral centers, defined in Section~\ref{sec:matched-motion}, are a distinct readout.
Each signed center uses every plaquette of its species:
$\bm R_z=\sum_\square\bm x_\square\phi_\square/(2\pi C[z])$.
Periodic coordinates require a chosen cut; sensitivity to that cut is checked.
Absolute density centers and covariance radii are used only for spatial shape diagnostics.
Table~\ref{tab:readouts} distinguishes these observables; their separations need not agree even on the same field.
\begin{table}[ht]
\caption{Center observables and preparations. All winding centers include the complete signed density of their species; none is a constituent position imposed in minimization.}
\label{tab:readouts}
\begin{ruledtabular}
\begin{tabular}{p{0.24\textwidth}p{0.40\textwidth}p{0.27\textwidth}}
Observable & Definition & Use\\
Moment spectral centers & Signed plaquette centers of the highest and lowest eigenlines of $M=pR_v-qR_w$ & Matched releases\\
Raw marginal centers & Signed plaquette centers of the principal eigenlines of $R_v,R_w$ & Independent readout control\\
Absolute density centers & Centers with weights $|b_v|,|b_w|$; the $v$ density can be partitioned along its covariance axis & Shape, radii, and the two outer trimer lobes\\
Constituent depletion centers & Circular centers of $1-|v_1|^2$ and $1-|w_3|^2$ from the reconstructed frame & ``Cloud separation'' in moving pair continuation\\
Field center & Circular center of $\langle\bar Q\rangle$ & Independent velocity fit; not used to fit the translated profile\\
\end{tabular}
\end{ruledtabular}
\end{table}
\FloatBarrier
The common flag pair has signed line separation $13.52a$, increased to $14.02a$ for the matched release.
The independently refined periodic full $(1,1)$ rest state instead has moment spectral separation $13.50a$ and depletion separation $12.28a$.
Thus the quoted $14.02$ and $12.28$ refer both to different preparations and to different observables.
The first is a perturbed flag frame used identically across representations; the second belongs to the full rest state from which the neutral moving family is constructed.

For normalized local rays the projective distance is
\begin{equation}
 d^2(\psi,\varphi)=\sum_i\left(1-|\psi_i^\dagger\varphi_i|^2\right) \,.
 \label{eq:distance}
\end{equation}
Small differences are evaluated with a stable, locally phase aligned formula.
Unless explicitly stated, spatial translations and global Cartan rotations are not fitted.
Rays of different dimensions are never directly compared; a flag frame is lifted into the appropriate representation first.

The product action and its full horizontal TDVP are
\begin{equation}
 S=\int\dd\tau\left[\ii\sum_i\psi_i^\dagger\dot\psi_i-E\right] \,,\qquad
 \Omega(u,v)=2\operatorname{Im}\sum_i u_i^\dagger v_i \,,\qquad
 \ii\dot\psi_i=P_i^\perp G_i \,.
 \label{eq:TDVP}
\end{equation}
Flag evolution instead uses the pullback of this form, with weights $(p,q)$.
The full evolution uses DOP853 at fixed time steps and does not renormalize or impose charge corrections during evolution.
Independent canonical implicit midpoint controls test the same normalized ray energy.
During implicit midpoint stages the intermediate vectors need not have unit norm. Dividing the horizontal right hand side by $n_i=\psi_i^\dagger\psi_i$ gives the canonical extension that preserves the Cartan invariants for these intermediate fields.
Energy, raw norms, raw and normalized Cartan expectations, topology, and complete ray errors are checked separately.

\section{Unrestricted pair minima and trimer robustness}
\label{sec:statics}\label{sec:static}
We first test whether both compositions survive when every local amplitude is relaxed, then compare their energies with separately optimized fragments.
The extended attraction and stiffness scans concern the neutral $(1,2)$ trimer.
The other composition--representation combinations provide the reference comparison for the change in neutrality.
\subsection{Diagonal mixed pair in both representations}
The mixed class $\bm k=(-1,+1)$ has $(C_v,C_w)=(-1,-1)$.
Its ambient degree is zero for $(1,1)$ and $+1$ for $(1,2)$.
A pair aligned with a lattice axis is a rotational saddle: after stationary refinement its lowest curvature is
$-2.006\times10^{-6}$ in $\mathbb{CP}^{7}$ and $-1.520\times10^{-6}$ in $\mathbb{CP}^{14}$.
The axial pair restricted to the flag likewise has a resolved negative curvature $-1.516\times10^{-6}$.
Relaxing along this unstable orientation gives the diagonal pair branch in Table~\ref{tab:pairs}.

\begin{table}[ht]
\caption{Unrestricted diagonal pair minima at $\eta/h=4$, with a two site vacuum collar. The comparison uses separately optimized $N=129$ root fragments, $E_A+E_B=0.9749699$; it is not a global fragment sector infimum.}
\label{tab:pairs}
\begin{ruledtabular}
\begin{tabular}{ccrrr}
Local space & $N$ & Energy & $\max_i\|\nabla_iE\|$ & $E_A+E_B-E_{\rm pair}$\\
$\mathbb{CP}^{7}$ &65&0.9733951&$1.51\times10^{-14}$&0.001575\\
$\mathbb{CP}^{14}$&65&0.9733993&$1.77\times10^{-12}$&0.001571\\
$\mathbb{CP}^{7}$ &97&0.9733858&$6.80\times10^{-14}$&0.001584\\
$\mathbb{CP}^{14}$&97&0.9733900&$4.58\times10^{-14}$&0.001580\\
\end{tabular}
\end{ruledtabular}
\end{table}
The $N=65$ to 97 energy change is about $9.34\times10^{-6}$, below $0.6\%$ of either tested binding margin.
All eight precisely evaluated low eigenpairs in each $N=97$ full space are positive, with residuals below $1.85\times10^{-12}$.
Table~\ref{tab:lowmodes} compares them with the neutral trimer.
The first two pair curvatures describe weak finite collar translation pinning.
The lower bounds for the normal principal block are $8.743$ and $7.662$ in the two pair spaces.
Their maximum local distances to a polar flag reconstruction are $1.870\times10^{-4}$ and $1.836\times10^{-4}$.
That reconstruction raises their energies by $1.479\times10^{-5}$ and $1.056\times10^{-5}$; these are projection costs, not differences from independently optimized flag minima.

\begin{table}[ht]
\caption{Eight full low curvatures at $N=97$, $\eta/h=4$, after removing only the two exact Cartan phase directions. All three fields have a vacuum collar. The first two are translations; the third is the orientation mode.}
\label{tab:lowmodes}\label{tab:lowcurvatures}
\begin{ruledtabular}
\begin{tabular}{crrr}
Mode & Neutral $\mathbb{CP}^{7}$ pair & Charged $\mathbb{CP}^{14}$ pair & Neutral $\mathbb{CP}^{14}$ trimer\\
1&$1.569\times10^{-10}$&$1.046\times10^{-10}$&$5.093\times10^{-10}$\\
2&$1.570\times10^{-10}$&$1.047\times10^{-10}$&$5.108\times10^{-10}$\\
3&$1.912\times10^{-6}$&$1.439\times10^{-6}$&$2.050\times10^{-6}$\\
4&$1.504\times10^{-4}$&$1.101\times10^{-4}$&$8.207\times10^{-5}$\\
5&$1.237\times10^{-3}$&$1.073\times10^{-3}$&$2.461\times10^{-4}$\\
6&$1.360\times10^{-3}$&$1.170\times10^{-3}$&$1.272\times10^{-3}$\\
7&$1.716\times10^{-3}$&$1.713\times10^{-3}$&$1.677\times10^{-3}$\\
8&$1.756\times10^{-3}$&$1.754\times10^{-3}$&$1.685\times10^{-3}$\\
\end{tabular}
\end{ruledtabular}
\end{table}

\subsection{Neutral trimer, attraction scan, and independent controls}
The neutral trimer has no topological protection in the full state space, so its survival depends on energy costs.
We compare it with three possible breakup configurations and follow its lowest deformations as the attraction changes.
For the $(1,2)$ trimer, $(C_v,C_w)=(-2,-1)$ and $C_{\rm full}=0$.
In the $\bm k$ convention, $A=(-1,0)$ and $B=(0,+1)$ are the two species; $D=A_2=(-2,0)$ is a dimer and $M=AB=(-1,+1)$ is a mixed pair.
The tested sign compatible partitions of $2A+B$ are $D+B$, $M+A$, and $A+A+B$.
At the reference parameters the independently optimized $N=129$ root energies are
\begin{equation}
 E_A=0.2870124 \,,\qquad E_B=0.6879576 \,,\qquad E_D=0.5334109 \,.
 \label{eq:fragments}
\end{equation}
Their changes from $N=97$ are $7.12\times10^{-10}$, $1.08\times10^{-10}$, and $7.17\times10^{-9}$.
Independent endpoint evaluations give $E_D+E_B=1.2213685$ and $2E_A+E_B=1.2619823$, consistent with the reference sums before rounding.
The root $A,B$ fields were also lifted and relaxed independently in both representations at $N=97$; their energies agree to numerical precision.

The attraction annihilates these pure root rays, so their stationary energies are independent of $\eta$.
This does not alone settle transverse stability.
At $\eta/h=4$, the smallest checked nonzero transverse curvatures are $0.002526$ for $A$ on $N=81$, $0.002138$ for $D$ on $N=81$, and $0.0007868$ for $B$ on $N=65$.
For root $12$, the analysis includes all 26 local directions outside its $\mathbb{CP}^{1}$ family.
The root-$23$ zero color block retains the full $\mathbb{CP}^{2}$ tangent, including directions outside the embedded root sphere.
Numerical Hermitian LDL checks and complementary block bounds show no resolved transverse instability.
The values $-8.57\times10^{-11}$ in the dimer helicity block and $-3.21\times10^{-11}$ in the root-$23$ block are comparable to the numerical residuals, so their signs remain unresolved.
The perturbation from $\eta/h=4$ to 4.2 is bounded in Hessian norm by $2\Delta\eta\|K\|\le0.00032$, below the quoted checked nonzero transverse margin; $K$ annihilates the zero color blocks.
This carries those specific transverse checks across the attraction scan, at their checked box sizes.

\begin{table}[ht]
\caption{Ten point unrestricted neutral trimer continuation on $N=97$. Energies and gradients are independently reevaluated. The binding column compares with the measured $D+B$ sum and is evaluated before rounding.}
\label{tab:continuation}
\begin{ruledtabular}
\begin{tabular}{crrrrr}
$\eta/h$ & Energy & $D+B-E$ & $\lambda_{\rm rot}$ & $\lambda_{\rm shape}$ & $\max_i\|\nabla_iE\|$\\
3.990&1.2198832&0.001485&$2.059\times10^{-6}$&$8.450\times10^{-5}$&$1.3\times10^{-14}$\\
4.000&1.2192899&0.002079&$2.050\times10^{-6}$&$8.207\times10^{-5}$&$1.0\times10^{-14}$\\
4.025&1.2177933&0.003575&$2.026\times10^{-6}$&$7.584\times10^{-5}$&$1.7\times10^{-13}$\\
4.050&1.2162778&0.005091&$2.000\times10^{-6}$&$6.938\times10^{-5}$&$3.7\times10^{-13}$\\
4.075&1.2147433&0.006625&$1.971\times10^{-6}$&$6.270\times10^{-5}$&$1.3\times10^{-12}$\\
4.100&1.2131901&0.008178&$1.941\times10^{-6}$&$5.581\times10^{-5}$&$1.4\times10^{-12}$\\
4.125&1.2116182&0.009750&$1.908\times10^{-6}$&$4.872\times10^{-5}$&$2.9\times10^{-14}$\\
4.150&1.2100278&0.01134&$1.873\times10^{-6}$&$4.143\times10^{-5}$&$1.2\times10^{-14}$\\
4.175&1.2084190&0.01295&$1.836\times10^{-6}$&$3.396\times10^{-5}$&$1.1\times10^{-14}$\\
4.200&1.2067918&0.01458&$1.797\times10^{-6}$&$2.632\times10^{-5}$&$4.8\times10^{-14}$\\
\end{tabular}
\end{ruledtabular}
\end{table}
All eight computed low curvatures at every point are positive; the largest of the 80 full eigenpair residuals is $1.69\times10^{-12}$.
The largest independent site gradient is $1.41\times10^{-12}$.
Adjacent first shape mode overlaps exceed $0.9983$, orientation overlaps exceed $0.9978$, and the minimum translation subspace overlap singular value exceeds $0.9992$.
Energy secants lie between their endpoint Hellmann--Feynman slopes $-h\sum_i\langle K_i\rangle$.
The first shape mode softens significantly toward 4.2, so no extension beyond the sampled window is claimed.
A zero obtained by fitting or extrapolating these positive curvatures would not locate a branch endpoint or identify the configuration reached after instability.
The choice $\eta=4h$ is a convenient reference also covered by the elementary bound for general representations, not a physical transition: the stronger common vacuum gap bound includes the complete scan.

Both attraction endpoints were enlarged to $N=129$ and refined without a center constraint.
Their energy changes are $-4.348\times10^{-8}$ and $-3.690\times10^{-8}$; the relative changes in first shape curvature are $2.584\times10^{-6}$ and $1.544\times10^{-5}$.
The corresponding orientation changes are below $2.5\times10^{-4}$ relatively.
Translation curvatures decrease to order $1.8\times10^{-13}$, with precise residuals near $5$--$6\times10^{-15}$.
They are finite collar quantities, not a bulk pinning gap or a measured Peierls barrier.
Independent searches for the smallest algebraic eigenvalues corroborate the nontranslational spectrum.

At $\eta/h=4.1$, changing $\rho$ by $\pm5\%$ and separately relaxing the pure fragments again gives Table~\ref{tab:rho}.
On this stiffness axis, the pure root fragments are independently relaxed, but the mixed pair and complete fragment transverse spectra are not recomputed.
\begin{table}[ht]
\caption{Independent stiffness control at $\eta/h=4.1$. The trimer has positive computed low spectra at all three values.}
\label{tab:rho}
\begin{ruledtabular}
\begin{tabular}{crrr}
$\rho$ & Trimer energy & Recomputed $D+B-E$ & $\lambda_{\rm shape}$\\
0.19&1.2019064&0.008760&$5.416\times10^{-5}$\\
0.20&1.2131901&0.008178&$5.581\times10^{-5}$\\
0.21&1.2244653&0.007600&$5.747\times10^{-5}$\\
\end{tabular}
\end{ruledtabular}
\end{table}

The diagonal mixed pair $M=AB$ was continued to both attraction endpoints to check the $M+A$ breakup comparison.
Its $N=97$ energies are $0.9734664$ at 3.99 and $0.9714388$ at 4.20, with gradients near $10^{-14}$ and positive precise low spectra corroborated by independent searches.
The $M+A$ margins above the trimer are $0.04060$ and $0.05166$.
Thus, $D+B$ remains the lowest tested breakup comparison.

The positive energy differences compare the trimer with specified relaxed fragments.
They measure binding against those comparisons, without imposing matching Cartan charges on the fragments.
A decay threshold for a closed system would require that additional constraint, and an escape barrier requires a continuous path rather than only endpoint energies.
The path calculation below addresses the latter question; Section~\ref{sec:data} states the remaining limits of the stability and binding tests.

\subsection{Independent change of the collapse stiffness}
\label{sec:alpha-control}
Changing the attraction alone could conceal a finely tuned balance with the cost of collapse.
We therefore change that stiffness independently and recompute both the molecule and its fragment comparison.
The coefficient $\alpha$ controls both the Laplacian cost and the local purity coefficient $s_0=8\rho+40\alpha$.
We change it by $\pm10\%$ while holding $\rho,h,w_F,w_B$ fixed, and refine the unrestricted neutral trimer at $\eta/h=3.99,4.00,4.20$.
The pure root $D$ and $B$ references are separately refined at each $\alpha$; their sums are $1.1689279E_0$ and $1.2712287E_0$, respectively.
Direct checks confirm that the attraction annihilates these reference fields.
All six trimers in Table~\ref{tab:alpha-control} have $C_{\rm full}=0$, constituent degrees $(-2,-1)$, maximum site gradients below $3.9\times10^{-13}$, and eight positive computed low curvatures after removal of the two Cartan phase directions.
Independent full Hessian residuals are below $1.6\times10^{-11}$. Separate random searches for the smallest algebraic eigenvalues at $\eta/h=4$ recover all eight modes at both additional $\alpha$ values, with residuals below $1.8\times10^{-10}$.
\begin{table}[ht]
\caption{Independent $\alpha$ controls on $N=97$ with a two layer vacuum collar. The shape curvature excludes translations and orientation. Energies are in $E_0$; curvatures use the real tangent metric of Section~\ref{sec:methods}.}
\label{tab:alpha-control}
\begin{ruledtabular}
\begin{tabular}{ccrrrr}
$\alpha$ & $\eta/h$ & Trimer energy & $D+B-E$ & $10^6\lambda_{\rm rot}$ & $10^5\lambda_{\rm shape}$\\
3.6&3.99&1.1680689&0.0008591&2.200&8.602\\
3.6&4.00&1.1675091&0.001419&2.190&8.362\\
3.6&4.20&1.1557177&0.01321&1.927&2.831\\
4.4&3.99&1.2691444&0.002084&1.941&8.319\\
4.4&4.00&1.2685192&0.002709&1.932&8.075\\
4.4&4.20&1.2553490&0.01588&1.689&2.462\\
\end{tabular}
\end{ruledtabular}
\end{table}
\FloatBarrier
Thus binding relative to the tested pure fragments and positive computed low curvatures survive at all three attraction values on both additional stiffness slices. The comparison uses stationary pure fragments; their complete transverse spectra are not recalculated on this axis.

\subsection{Explicit unconstrained escape paths}
\label{sec:escape-paths}

We construct continuous paths from the neutral $(1,2)$ trimer to the
polarized vacuum at $\eta=4h$.  These provide numerical upper bounds on an
unconstrained finite lattice escape barrier.  They do not establish a
barrier lower bound, a minimum energy saddle, or a lifetime.  The primary
path uses the same $97\times97$ lattice and fixed two layer vacuum collar
as the reference state.  Its first image is the exact field paired with
the reference Hessian spectrum.  A short connecting segment, whose largest
local ray displacement is below $7\times10^{-8}$, joins this field to the
independently polished state used to initialize the string.

For consecutive normalized image rays $a_i,b_i$, choose the phase of
$b_i$ so that $c_i=\langle a_i|b_i\rangle\geq0$, and define
\begin{equation}
 \theta_i=\arccos c_i\,,\qquad
 u_i=\frac{b_i-c_i a_i}{\sin\theta_i}\,,\qquad
 \psi_i(t)=a_i\cos(t\theta_i)+u_i\sin(t\theta_i)\,,
 \quad(0\leq t\leq1)\,.
 \label{eq:escape-geodesic}
\end{equation}
Coincident rays give constant segments.  Thus every pair of stored images
is joined explicitly, including the reference state connector.  For any
local operator, its expectation on a segment is exactly of the form
$C_i+D_i\cos(2\theta_i t)+S_i\sin(2\theta_i t)$.
Substitution into the actual finite range bond energy, including the
vacuum onsite subtraction, gives a finite trigonometric expansion
\begin{equation}
 E(t)=E_{\mathrm c}+\sum_k
 \bigl[r_k\cos(\omega_k t)+s_k\sin(\omega_k t)\bigr]\,,\qquad
 |E'''(t)|\leq M_3\equiv
 \sum_k|\omega_k|^3\sqrt{r_k^2+s_k^2}\,.
 \label{eq:escape-third-bound}
\end{equation}
The frequencies include $2\theta_i$ and $2\theta_i\pm2\theta_j$.
On an interval of midpoint $m$ and half width $\delta$,
\begin{equation}
 \max E(t)\leq
 \max_{|s|\leq\delta}
 \left[E(m)+E'(m)s+\frac{E''(m)}2s^2\right]
 +\frac{M_3\delta^3}{6}\,.
 \label{eq:escape-energy-bound}
\end{equation}
The quadratic maximum is evaluated at both endpoints and any interior
maximum.  Adaptive subdivision reduces the global enclosure gap below
$10^{-5}E_0$.  This bounds the energy between samples and images.  Direct
bond contractions and the harmonic evaluation agree to approximately
$5\times10^{-12}E_0$.  The floating point calculation includes a
$10^{-8}E_0$ evaluation allowance; it is not an interval arithmetic
certificate.  Table~\ref{tab:escape-bounds} reports coarser, conservative rounded bounds.

\begin{table}[t]
\caption{Explicit trimer to vacuum paths.  Both include the exact reference
connector and start at $E/E_0\simeq1.2192899$.  The bounds concern the
recorded paths, not an optimized saddle.}
\label{tab:escape-bounds}
\begin{ruledtabular}
\begin{tabular}{lccc}
Path & $\max E/E_0$ upper & $\Delta E/E_0$ upper & Regular interpolation\\
Guarded string & $28.747$ & $27.53$ & Yes\\
Unguarded string & $6.585$ & $5.366$ & No\\
\end{tabular}
\end{ruledtabular}
\end{table}

The initial contraction was shortened by unrestricted horizontal gradient
steps and ray arclength redistribution.  The guarded attempt stopped when
the spatial guard prevented a further accepted step; no string convergence
is assumed.  Its continuous spatial regularity is checked separately.
For a neighboring overlap $z_{ij}$,
$|z'_{ij}|\leq\theta_i+\theta_j\leq2\max_i\theta_i$.
Sampling each segment at 65 parameter values and applying this Lipschitz
bound gives $|z_{ij}|\geq0.8134>\cos(\pi/4)$ throughout.
For a fixed diagonal triangulation, both nearest neighbor vertices of
each triangle therefore lie inside a strongly convex projective ball of
radius below $\pi/4$ around its anchor.  Shared diagonals have length below
$\pi/2$ and unique short geodesics, giving compatible continuous fillings.
The separately bounded absolute plaquette phase is below $0.3663<\pi$;
the ambient degree remains zero.  This supplies a regular spatial
interpolation of the lattice homotopy, not an energy bound for an
interpolated continuum functional.  The lower energy unguarded path crosses
plaquette branches and temporarily has lattice degree $-2$.

The two total Cartan expectation sums change by approximately $435.0$
and $164.4$ between endpoints.  Neither path is a charge preserving
closed system decay trajectory.  A separate direct construction toward
compact D and B fragments 52 sites apart on a $129\times129$ lattice was
also retained, but its loose lattice bound and failed regularity check do
not qualify an informative dissociation barrier.  A regular fission path
or connecting saddle remains uncomputed.

\subsection{The charged trimer in the eight state representation}
\label{sec:reciprocal-trimer}
To test the reciprocal representation comparison, the optimized flag trimer in $(1,2)$ at $\eta/h=4$, $N=97$, is lifted into $(1,1)$ and released in all local $\mathbb{CP}^7$ directions with the same two layer vacuum collar.
The common flag energy is approximately $1.219372$ in both lifts.
Stationary refinement gives an unrestricted state with independently evaluated energy approximately $1.219259$, maximum full gradient $4.98\times10^{-13}$, and charges $(C_v,C_w)=(-2,-1)$, $C_{\rm full}=-1$.
The independent evaluation contracts the physical bond operators and horizontal gradient directly, without importing the minimizer's energy or gradient routine.

Both exact Cartan phase directions have resolved rank; no position or shape symmetry is imposed in the final stationarity or spectra.
The eight precise low curvatures are
\begin{equation}
 \begin{split}
 \lambda={}&(6.842\times10^{-10},\,6.926\times10^{-10},\,
 2.432\times10^{-6},\,1.042\times10^{-4},\\
 &3.098\times10^{-4},\,1.457\times10^{-3},\,
 1.712\times10^{-3},\,1.745\times10^{-3}) \,.
 \end{split}
 \label{eq:cp7-trimer-curvatures}
\end{equation}
Their full residuals are below $1.64\times10^{-12}$; an independent random initialization recovers all eight with full residuals below $1.88\times10^{-10}$.
The first two modes are weak collar translations and the third is orientation.
A Gershgorin estimate bounds the normal principal block below by $8.750$; coupling to the other directions is included in the full Hessian searches.
The gap between the two largest eigenvalues of each constituent density matrix exceeds $1-1.9\times10^{-7}$, and the maximum local distance to the polar flag reconstruction is $3.727\times10^{-4}$.
The gradient and two spectral searches meet the same numerical local stability criteria as the other three states.

The optimized $N=129$ pure root $D=A_2$ and $B$ fields are lifted into $(1,1)$ and independently evaluated in its full local space.
A spatial reflection puts $B$ in the required $C_w=-1$ orientation without changing its square lattice energy.
Their energies are $0.5334109$ and $0.6879576$, with maximum full gradients below $3.83\times10^{-13}$.
Using the unrounded energies, the tested $D+B$ margin is $0.002110$.
These are stationary lifted reference fragments; their full $(1,1)$ transverse spectra and global constrained fragment infima have not been newly established.

Together with the pair minima and neutral $(1,2)$ trimer, this supplies all four static composition--representation combinations.
The charged trimer is tested only at this reference point, without an attraction scan or a larger box check. Its reciprocal dynamical release is reported in Section~\ref{sec:trimer-release}.
Its nonzero ambient degree also distinguishes it from the neutral classes that can contract smoothly in the full local space.

\subsection{Energy mechanism, shape, and periodic rest controls}
The energy decomposition shows what holds the neutral trimer together and how closely its full state remains confined to the flag.
The shape measurements and periodic checks then separate the molecular size from effects of the frozen boundary.
The rounded reference field energy contributions are
\begin{equation}
 \begin{array}{c|rrrrr}
 &\text{gradient}&\text{Laplacian}&\text{purity deficit}&\text{field}&\text{attraction}\\ \hline
 E&0.2252&0.5043&0.00008202&0.7276&-0.2379
 \end{array} \,.
 \label{eq:energybalance}
\end{equation}
The attraction rewards overlap of the two constituent deviations, while the gradient, Laplacian, and onsite field penalize spatial structure.
The positive purity cost confines the state near the flag without imposing it.
The attraction energy is much larger than the final dissociation margin; comparing separately optimized states at different $\eta$ would not isolate this term.
For the fixed reference ray, $\mathcal K_*=\sum_i\langle K_i\rangle\simeq74.35$ and
\begin{equation}
 E_\eta[\psi_*]=E_{4h}[\psi_*]+(4h-\eta)\mathcal K_* \,.
 \label{eq:reference-affine-energy}
\end{equation}
Relative to the fixed pure root $D+B$ fields, this trial energy is lower for $\eta/h>3.966$.
This affine ordering is not a stable branch interval and is not used in place of Table~\ref{tab:continuation}.
At $\eta=0$ the separately relaxed candidate has energy $1.2214078$, whereas an explicit disjoint $D+B$ trial in one $N=225$ box has energy $1.2213698$. The candidate lies higher by $3.80\times10^{-5}$ and therefore does not establish binding there.

In Figure~2 of the Letter, the displayed sign convention is $\nu_A=-q_v$ and $\nu_B=-q_w$, where $q_z=\phi_\square[z]/(2\pi)$ is the winding per plaquette.
The plotted binding is $\Delta E_{\rm bind}=E_{A_2}+E_B-E_{A_2B}$, with $D=A_2$ in the tables here.
The diagonal trimer's root-$12$ absolute density, divided through its covariance axis with fractional plaquette areas, has outer centers $(\mp5.726,\mp5.726)$ and outer separation $16.20$.
The central root-$23$ center is within $4\times10^{-7}$ of the origin.
Outer to central distances are $8.098$; RMS charge radii are $8.074,8.074,5.503$.
The lobes overlap appreciably.
These absolute density diagnostics are distinct from signed Berry centers and are not minimization constraints.

After smooth random perturbations and perturbations along both signs of the first shape mode, eight full relaxations on $N=81$ return within $1.3\times10^{-13}$ in energy and below $2\times10^{-9}$ in maximum gradient.
They test relaxation, not Hamiltonian evolution.
For periodic trimer rest states, $E_{97}\simeq1.2192899$, with $E_{129}-E_{97}\simeq-7.93\times10^{-9}$.
The two translation curvatures on $N=97$ are around $-5\times10^{-15}$, smaller than the residuals of order $10^{-13}$, so their signs are unresolved.
The periodic orientation and first shape curvatures remain $2.050\times10^{-6}$ and $8.207\times10^{-5}$.
Relaxed center scans at fixed Cartan charges resolve no energy variation above the approximately $10^{-12}$ scatter at the sampled positions; this does not bound a barrier between them.
\section{Flag minima, full state relaxation, and topology}
\label{sec:coherent-bridge}
The common flag energy would have little predictive value if the texture changed drastically as soon as extra local directions were released.
We test this directly by minimizing on the flag and then relaxing the same Hamiltonian in the full state space.
A separate geometric construction explains why the integrated winding formula remains valid for the nearby full states.
\subsection{Energetic continuation in the same Hamiltonian}
The trimer comparison uses the same $N=97$ collar, diagonal branch, and Hamiltonian at each coupling.
A polar reconstruction is first independently minimized in every local flag direction.
The Hessian restricted to the flag is then evaluated as in Section~\ref{sec:methods}.
Releasing every $\mathbb{CP}^{14}$ direction under the unchanged Hamiltonian returns to the independently obtained unrestricted branch.
Table~\ref{tab:coherent-bridge} compares the flag minima with these unrestricted releases.

\begin{table}[ht]
\caption{Optimized flag trimers and unrestricted releases. The drop uses the full evaluator at both endpoints; vacuum reference offsets between energy formulas are below $10^{-10}$. Distances remove only local ray phases.}
\label{tab:coherent-bridge}
\begin{ruledtabular}
\begin{tabular}{crrrrr}
$\eta/h$ & $E_{\rm flag}$ & Relaxation drop & Final full gradient & Distance to full reference & $\lambda_{\rm shape}^{\rm coh}$\\
4.0&1.2193719&$8.203\times10^{-5}$&$1.51\times10^{-13}$&$9.42\times10^{-7}$&$8.223\times10^{-5}$\\
4.1&1.2132741&$8.404\times10^{-5}$&$5.68\times10^{-13}$&$7.91\times10^{-6}$&$5.599\times10^{-5}$\\
4.2&1.2068773&$8.552\times10^{-5}$&$1.17\times10^{-14}$&$5.51\times10^{-6}$&$2.651\times10^{-5}$\\
\end{tabular}
\end{ruledtabular}
\end{table}
The maximum site gradients within the flag are $1.6\times10^{-13}$, $3.3\times10^{-13}$, and $8.1\times10^{-15}$.
The complete six direction spectra, after exact Cartan phases are removed, have resolved positive internal and orientation modes.
The orientation curvatures are $2.05\times10^{-6}$, $1.94\times10^{-6}$, and $1.80\times10^{-6}$.
Released energies agree with full references within $2.5\times10^{-13}$.
The optimized flag states still have ambient gradients of maximum local norm approximately $0.0035$, directed normally to the flag orbit.
The full minima differ by maximum local normal displacements $3.54$--$3.62\times10^{-4}$.
Thus, the ansatz is accurate but not exact; the same soft internal mode is present on both stationary branches.

For the mixed pair, the common diagonal flag minimum has $E=0.9734099$ on $N=65$ and $E=0.9734005$ on $N=97$ with a collar.
The latter has maximum flag gradient $5.76\times10^{-13}$.
Releasing the collar creates a boundary force of order $1.4\times10^{-5}$, so the dynamical common frame is separately refined with all sites periodic.
It has $E_{\rm flag}\simeq0.9734005$, a decrease of $7.13\times10^{-9}$ from the collar minimum, gradient near $10^{-14}$, and $(C_v,C_w)=(-1,-1)$.
Its orientation and first shape curvatures in the $(1,2)$ induced metric are $1.435\times10^{-6}$ and $1.096\times10^{-4}$.
Translation signs around $10^{-14}$ remain unresolved.
This same periodic flag frame is used for both representations in Section~\ref{sec:matched-motion}.
Its lifts of the flag are not ambient stationary states.

\subsection{Exact homotopy near the flag}
The full minima lie close to flag configurations, but proximity alone does not establish equality of their topological charges.
The following explicit deformation connects a full field to its reconstructed flag whenever the stated eigenvalue and overlap conditions hold.
For a smooth full projective field $f$, suppose both marginal principal eigenvalues are isolated.
Choose local normalized principal eigenvectors and form $Z=(v,w)$.
If $Z^\dagger Z$ is positive definite, the symmetric polar frame
$\widetilde Z=Z(Z^\dagger Z)^{-1/2}$ defines a smooth flag independently of eigenvector phases.
Let $g=\iota_{p,q}(\widetilde Z)$ be its coherent lift.
If the pointwise ray overlap never vanishes, a based projective homotopy is
\begin{equation}
 H_t(f)=\left[(1-t)\psi+
 t\,\frac{P_g\psi}{\sqrt{\langle\psi|P_g|\psi\rangle}}\right] \,,
 \qquad 0\le t\le1 \,,
 \label{eq:homotopy}
\end{equation}
where $\psi$ is any local normalized lift of $f$.
The two terms have positive real overlap, so their sum does not vanish.
The construction is phase covariant and fixes a vacuum boundary.
The raw principal lines are themselves homotopic to their polar corrected lines through
$Z(Z^\dagger Z)^{-s/2}$.
Consequently the integrated identity
\begin{equation}
 C_{\rm full}[f]=pC_v-qC_w
 \label{eq:nearcoherent-topology}
\end{equation}
is exact throughout this neighborhood, although the local Berry density decomposition is exact only on the flag orbit.
An unwinding path can leave the neighborhood; the retraction supplies no global topological protection or escape barrier.
Equation~\eqref{eq:homotopy} connects a full ray field to its nearby flag reconstruction, not to the vacuum.
Neutral coherent classes admit a smooth ambient contraction by Section~\ref{sec:geometry}, whereas a charged class with nonzero full degree cannot be smoothly contracted to the vacuum with a fixed boundary.
On a discrete lattice a collapse path may instead leave the admissible smooth regime; this is a distinct mechanism.

For the reference trimer, principal gaps exceed $1-2.0\times10^{-7}$ and $1-1.0\times10^{-7}$,
the minimum frame Gram eigenvalue differs from unity by $2.216\times10^{-4}$, and the minimum coherent overlap amplitude has a deficit $6.77\times10^{-8}$.
The maximum phase aligned local displacement obeys $\epsilon<3.680\times10^{-4}$.
The minimum neighboring ray overlap amplitude obeys $r>0.9617$, and the maximum plaquette phase is below $0.1522$.
Along the normalized chordal homotopy, a link changes by at most $2\epsilon$; a plaquette phase stays below
$0.1522+4\arcsin(2\epsilon/r)<0.1553<\pi$.
The raw to polar line interpolations have still smaller margins and preserve $(C_v,C_w)=(-2,-1)$.
This is a checked admissible discrete continuation for the stored field, not a certificate for every possible interpolation between lattice sites.

\section{Matched motion and the moment map bridge}
\label{sec:matched-motion}
These releases compare like with like: one flag texture receives the same geometric deformation and is then evolved in each representation.
The energy at launch is identical, while the Berry weights differ.
We first identify a common observable of motion, then compare the complete trajectories with evolution restricted to the flag.
\subsection{An exact initial bridge through the physical moment}
A fair comparison of full and flag motion needs an observable with the same initial response in both descriptions.
The total $SU(3)$ moment has this property, even though the raw constituent density matrices can begin to change in different ways.
Besides the raw marginal lines of Equation~\eqref{eq:marginals}, consider the exact total $SU(3)$ moment matrix
\begin{equation}
 {\cal M}=pR_v-qR_w
 =2\sum_a\langle T^a\rangle t^a+\frac{p-q}{3}I \,.
 \label{eq:fullmoment}
\end{equation}
Its highest and lowest spectral eigenlines define the moment spectral centers.
On the flag orbit $\mathcal M=pP_v-qP_w$, with eigenvalues $(p,0,-q)$;
the traceless matrix in Equation~\eqref{eq:flagmoment} is $\mathcal M-(p-q)I/3$.
Off the orbit the spectral lines and the two raw marginal principal lines are distinct observables.

The coherent embedding is Kähler.
At an initial flag field, the tangent space and its metric orthogonal normal complement are each invariant under the ambient complex structure.
The tangent part of the full Hamiltonian vector field is exactly the Hamiltonian vector field of the restricted flag energy.
Every local $SU(3)$ moment generates an orbit tangent vector field, so its differential annihilates the symplectic normal component.
For the same initial flag configuration and Hamiltonian,
\begin{equation}
 \left.\dot{\cal M}_{\rm full}\right|_{\rm coh}
 =\dot{\cal M}_{\rm coh} \,.
 \label{eq:moment-bridge}
\end{equation}
The identity holds at launch. Subsequent full motion can leave the flag, so it need not coincide with the restricted evolution.

Direct tensor marginal differentiation on the actual perturbed preparation verifies Equation~\eqref{eq:moment-bridge} with relative errors $2.05\times10^{-11}$ for $q=1$ and $4.04\times10^{-9}$ for $q=2$.
Independent center probes at derivative steps $10^{-3}$ and $5\times10^{-4}$ agree within approximately $2.2\times10^{-11}$ in velocity.
Normal motion can alter the raw marginal eigenlines while its contributions cancel in $p\,\delta R_v-q\,\delta R_w$.
Their initial raw center midpoint speeds are about 1.91 and 1.75 times the flag values, while the moment spectral initial velocities agree with the flag ones.
The difference concerns the choice of center observable; it is consistent with the common TDVP normalization.

For a rigid two center continuum approximation with interaction depending on separation,
equal and opposite forces act on constituent gyrocharges $-2\pi n_0p$ and $+2\pi n_0q$.
Their transverse displacement magnitudes therefore have ratio $q:p$.
An explicit four coordinate projector shift pullback, including the measured map from input shifts to signed centers, predicts initial flag center speeds to $1.56\%$ for $q=1$ and $9.07\%$ for $q=2$.
Its full field residual is nevertheless about $0.99$ of the field velocity.
Four translations capture a center response but not the complete dynamics; the small residual of their four by four force solve is not a rigid motion test.

\subsection{Identical preparation and completed trajectories}
Start with the common periodic flag minimum described above.
Translate its two line projectors oppositely along their separation axis by Fourier interpolation, extract their principal lines, and orthogonalize symmetrically by polar decomposition.
A requested relative increment $0.5$ produces a measured signed center increment $0.5003$, from $13.52$ to $14.02$.
The resulting single frame is used for both Berry weights and both full lifts of the flag.
Initial flag energies agree across representations; full/flag initial ray distances within the corresponding representation are below $7.1\times10^{-14}$.

Flag evolutions use steps $0.2$ and $0.1$; full evolutions use $0.1$ and $0.05$.
All eight runs reach $\tau=1000$, with 21 saved fields per run independently checked.
In Figure~1 of the Letter, symbols allow all local amplitudes to evolve and lines restrict evolution to flag states.
Table~\ref{tab:matched-motion} uses moment spectral centers for the full trajectories.
On the flag orbit this equals the ordinary line readout.

\begin{table}[ht]
\caption{Matched pair motion at $\tau=1000$. Travel is displacement along the initial transverse direction $(1,-1)/\sqrt2$, in lattice units. Full trajectories use moment spectral centers. Relative angle changes are in radians, with counterclockwise rotation positive. Ratios are computed before rounding the travels; they are finite time responses, not exact lattice identities.}
\label{tab:matched-motion}
\begin{ruledtabular}
\begin{tabular}{ccrrrr}
Evolution & $(p,q)$ & $v$-line travel & $w$-line travel & Ratio & Relative angle change\\
Flag &$(1,1)$&0.047669&0.047689&0.99956&$-1.481\times10^{-6}$\\
Flag &$(1,2)$&0.054806&0.027426&1.9983&$+1.953\times10^{-3}$\\
Full &$(1,1)$&0.048346&0.048367&0.99956&$-1.508\times10^{-6}$\\
Full &$(1,2)$&0.055287&0.027667&1.9983&$+1.970\times10^{-3}$\\
\end{tabular}
\end{ruledtabular}
\end{table}
\FloatBarrier
The neutral constituents travel together, while the charged pair has relative turning.
In the full evolution, the charged angle is only $0.113^\circ$: the calculation resolves onset, not a completed orbit.
The full neutral trajectory has a maximum angular excursion over saved times of $6.91\times10^{-6}$ radians; its smaller endpoint angle is not a bound on preceding motion.
Travel is below $0.056$ lattice units, compared with separation $14.02$.

For the same full runs, raw marginal transverse travels are $(0.04834,0.04836)$ and $(0.05528,0.02766)$.
Their ratios are $0.99956$ and $1.9983$, and endpoint angles are $-1.520\times10^{-6}$ and $1.970\times10^{-3}$ radians.
Raw and moment centers differ by at most $1.61\times10^{-5}$.
Moment eigenvalue gaps exceed $1-9.0\times10^{-8}$ and eigenvalue departures from $(p,0,-q)$ stay below $8.7\times10^{-8}$.
The exact generator identity in Equation~\eqref{eq:fullmoment} holds within $1.9\times10^{-14}$.
Constituent and full charges are independently preserved, with the weighted degree identity accurate to $2.9\times10^{-15}$.
Full ray squared link overlaps exceed $0.9223$ and plaquette phases remain below $0.1629$.

\subsection{Integration, readout, and unperturbed release controls}
The displacements are small compared with the molecule, so their ratios must be distinguished from integration error and motion caused by the preparation itself.
Table~\ref{tab:matched-errors} gives the step comparison and conservation checks. An independent integration method also tests numerical accuracy; alternative center definitions and releases without the imposed stretch test the interpretation.
\begin{table}[ht]
\caption{Independent matched pair integration checks. Center differences use all saved moment spectral checkpoints; ray distances use Equation~\eqref{eq:distance}. Conservation columns refer to the fine full run.}
\label{tab:matched-errors}
\begin{ruledtabular}
\begin{tabular}{crrrrr}
$(p,q)$ & Step halving center error & Endpoint ray distance & Max $|\Delta E|$ & Max raw $|\Delta T^{3,8}|$ & Max norm$^2$ error\\
$(1,1)$&$1.56\times10^{-12}$&$1.34\times10^{-7}$&$5.56\times10^{-13}$&$6.46\times10^{-12}$&$1.83\times10^{-13}$\\
$(1,2)$&$5.87\times10^{-14}$&$1.14\times10^{-8}$&$4.83\times10^{-13}$&$4.61\times10^{-12}$&$2.09\times10^{-13}$\\
\end{tabular}
\end{ruledtabular}
\end{table}
\FloatBarrier
Raw center step differences remain below $8.9\times10^{-11}$; flag center differences are below $6.2\times10^{-14}$.
Full/flag moment center differences reach $4.86\times10^{-4}$ for $q=1$ and $3.48\times10^{-4}$ for $q=2$, about $1$--$2\%$ of the motion and well above integration error.
Coarse full energy changes are $3.48\times10^{-10}$ and $1.17\times10^{-11}$; both decrease at the finer step.
Independent canonical implicit midpoint on each actual preparation reaches $\tau=10$.
Distances to a fine DOP853 reference at steps $(0.025,0.0125,0.00625)$ are
$(2.243,0.5651,0.1415)\times10^{-4}$ for $q=1$ and
$(9.090,2.279,0.5701)\times10^{-5}$ for $q=2$, showing second order convergence.
This is a short independent method control, not a second full duration trajectory.

Absolute constituent charge in a two plaquette boundary strip is at most $4.87\times10^{-6}$.
Moving the periodic cut by four sites changes raw displacement coordinates by at most $5.18\times10^{-5}$ and full minus flag differences by at most $1.27\times10^{-6}$.
Weighted center numerators change by roughly $1.7$--$1.9\times10^{-4}$.
They are flag continuum diagnostics, not exact lattice momenta or unrestricted ray invariants.

Two unperturbed full releases use the same flag minimum with zero imposed relative displacement, at step $0.1$ through $\tau=1000$.
The lift is not ambient stationary, so these controls also contain normal readjustment.
Their moment spectral midpoint drifts are $0.0008173$ and $0.0004383$, only $1.69\%$ and $1.06\%$ of the perturbed drift.
Subtracting the measured rest control displacements gives midpoint travel $0.04762$ and $0.04105$, and angle changes $-1.491\times10^{-6}$ and $1.949\times10^{-3}$ radians.
These subtracted curves are comparison diagnostics, not separately evolved solutions.
Their endpoint species displacement ratios are $0.99956$ and $1.9983$, respectively.
The neutral/charged distinction survives.
Across all perturbed and unperturbed records, moment gaps exceed $1-1.1\times10^{-7}$.
The unperturbed initial velocities of the flag centers and full moment centers are unresolved near $5\times10^{-11}$, although raw marginal centers can have much larger rapidly varying initial derivatives.
Agreement of these center definitions and the releases without a stretch supports the interpretation of the perturbed response.
\subsection{Transient speed in the matched pair releases}
\label{sec:transient-speed}

\begin{figure}[ht]
\includegraphics[width=\textwidth]{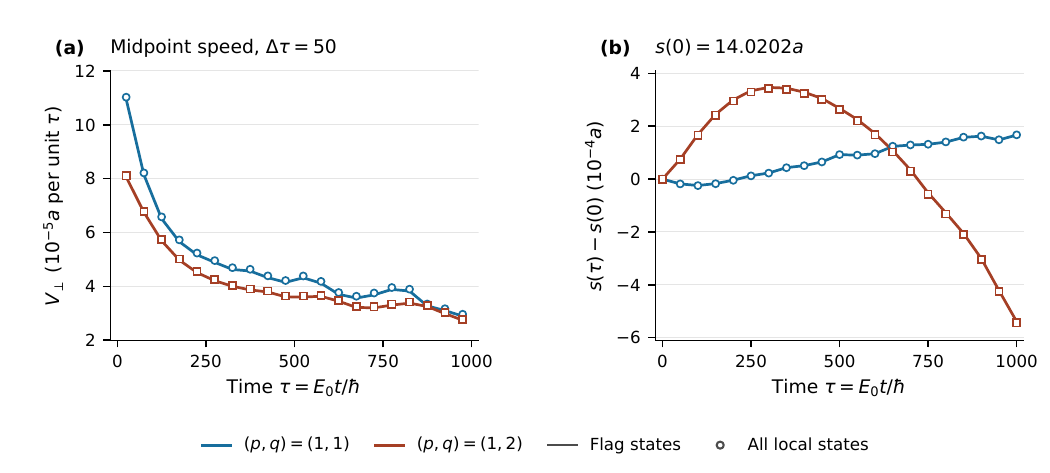}
\caption{Transient speed and nearly fixed separation in the matched pair releases.
(a) Transverse midpoint velocity $V_\perp$, evaluated as a secant over each saved interval of length $\Delta\tau=50$ and plotted at its midpoint.
(b) Change in separation of the two constituent centers from $s(0)=14.02a$.
Blue circles and red squares denote full space $(1,1)$ and $(1,2)$ data; lines denote the corresponding flag controls.
Centers are read from the spectral lines of the physical moment matrix, with flag eigenlines used in the controls.
The runs are periodic $97\times97$ releases with the same initial preparation.
Small changes in center separation do not measure all internal deformations.}
\label{fig:release-diagnostic}
\end{figure}

The matched preparations test the response to the same relative displacement; they are not constructed translating relative equilibria.
Figure~\ref{fig:release-diagnostic} shows the midpoint speed and constituent separation during these releases.
To quantify the curvature of the trajectories, we project the midpoint displacement perpendicular to the initial pair axis and form secants over each saved interval $\Delta\tau=50$.
The mean transverse speed over $500\leq\tau\leq1000$ is smaller than that over $0\leq\tau\leq500$ by $37.90\%$ for $(1,1)$ and $33.53\%$ for $(1,2)$ in the full spaces.
Comparing the first and last intervals of length $50$ instead gives $73.12\%$ and $65.72\%$; the stated slowdown therefore requires a specified time window.
The secant speeds have small reversals, so the trajectories are not strictly concave at every saved time.

The corresponding half interval decreases in flag dynamics are $38.35\%$ and $33.77\%$.
Subtracting the measured unperturbed release displacement from the full trajectories gives $38.43\%$ and $33.81\%$; this subtraction is a diagnostic difference between two solutions.
Thus, the slowing is already present within the flag manifold and cannot be assigned specifically to motion normal to it.
Meanwhile, the constituent separation, initially $14.02a$, changes by at most $1.68\times10^{-4}a$ and $5.45\times10^{-4}a$.
These changes are below $0.11\%$ of the imposed $0.5003a$ separation increment.
They do not describe appreciable relaxation of that increment, but do not exclude changing constituent shapes or other internal modes.

The continuum weighted first moment constrains relative displacements, not their time dependence.
Consequently the near $1{:}1$ and $2{:}1$ travel ratios are compatible with changing speeds.
The runs are periodic and contain no vacuum collar.
Energy and Cartan conservation and the time step controls remain satisfied.
The saved data establish a conservative transient response; they do not identify a radiation mechanism, an eventual speed, or a translating relative equilibrium.

\FloatBarrier

\subsection{Reciprocal matched trimer releases}
\label{sec:trimer-release}
In the symmetric $A$--$B$--$A$ trimer, the center of the two outer $A$ lobes coincides with the central $B$ center. The species centers therefore cannot define an axis before perturbation.
We instead use its molecular axis $(1,1)/\sqrt2$, shift the flag projectors for $v$ and $w$ by $+0.25a$ and $-0.25a$ along this axis, and reconstruct an orthogonal frame by principal eigenline extraction and symmetric polar orthogonalization.
The resulting species center separation is $0.5008a$.
Exactly this frame is lifted into both representations; its flag energy is identical in the two cases.
The source flag trimer was optimized with a vacuum collar, whereas these releases are periodic; the unperturbed controls below include the resulting initial readjustment.
All four full or flag releases run on periodic $N=97$ to $\tau=1000$.
No moving profile is fitted to these transients.

\begin{figure}[t]
\includegraphics[width=\textwidth]{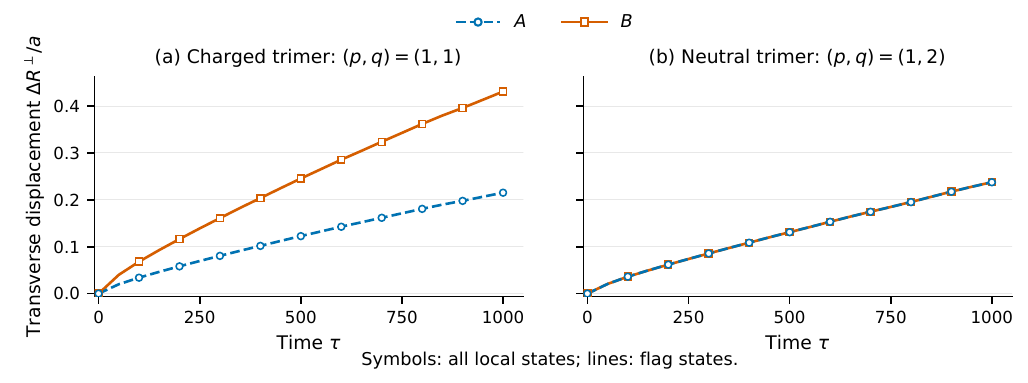}
\caption{Reciprocal matched trimer response to one common flag preparation. The full state symbols follow the flag curves: charged species travel at approximately a $1{:}2$ ratio, while neutral species travel together. The $A$ center includes both outer lobes. These are species center displacements, not whole molecule rotations or prepared rigid translations.}
\label{fig:trimer-response}
\end{figure}

For a flag field in the continuum, the signed species centers obey the first moment relation
\begin{equation}
 \frac{\bm D}{2\pi n_0}=-2p\bm R_v+q\bm R_w \,,\qquad
 2p\,\Delta\bm R_v=q\,\Delta\bm R_w \,.
 \label{eq:trimer-moment}
\end{equation}
Thus the predicted species displacement ratio is $q/(2p)$: $1/2$ for the charged $(1,1)$ trimer and unity for the neutral $(1,2)$ trimer.
Figure~\ref{fig:trimer-response} shows the reciprocal response in both full local spaces and its comparison with flag dynamics.
Table~\ref{tab:trimer-release} gives the endpoint displacements and the largest full/flag differences.
\begin{table}[ht]
\caption{Matched trimer release endpoints from the finer full state runs. The transverse direction is $(-1,1)/\sqrt2$. The last column is the largest difference between center coordinates in the full and flag trajectories over all saved checkpoints.}
\label{tab:trimer-release}
\begin{ruledtabular}
\begin{tabular}{ccrrrr}
$(p,q)$ & $C_{\rm full}$ & $\Delta R_v^\perp/a$ & $\Delta R_w^\perp/a$ & Ratio & Full/flag difference\,/$a$\\
$(1,1)$ & $-1$ & $0.21526$ & $0.43072$ & $0.49977$ & $2.69\times10^{-4}$\\
$(1,2)$ & $0$ & $0.23747$ & $0.23755$ & $0.99969$ & $1.31\times10^{-4}$\\
\end{tabular}
\end{ruledtabular}
\end{table}

The vector joining the species centers turns by $-0.4566$ radians in the charged case and $-1.455\times10^{-4}$ radians in the neutral case.
This is not a measurement of rigid rotation of the whole $A$--$B$--$A$ molecule: its initial species separation is only $0.50a$, much smaller than the outer lobe separation.
The principal axis of the absolute $v$ density changes by only $-0.002435$ and $-0.002662$ radians, respectively, modulo $\pi$.
The reciprocal test therefore establishes the species response associated with changed Berry weights, not a switch to rigid whole trimer rotation.

Full releases at steps $0.1$ and $0.05$ agree in moment center coordinates to $1.6\times10^{-12}a$ for $(1,1)$ and $8.7\times10^{-14}a$ for $(1,2)$.
The corresponding endpoint projector Hilbert--Schmidt distances, summed over the whole lattice and not normalized by its area, are $2.92\times10^{-7}$ and $2.47\times10^{-8}$.
The finer full energy drifts are below $2.4\times10^{-12}E_0$, Cartan drifts below $4.1\times10^{-11}$, and local norm errors below $1.9\times10^{-13}$.
Flag controls use step $0.1$; their energy drift is below $8\times10^{-15}E_0$.
Both unperturbed full controls have center displacement coordinates below $7\times10^{-8}a$. At their coarser step $0.1$, energy drifts are below $1.2\times10^{-9}E_0$ and local norm errors below $6.7\times10^{-13}$; the tighter bounds above apply to the finer perturbed runs.
Their centers coincide initially to about $10^{-12}a$, so their relative angle and displacement ratio are undefined and are not subtracted from the perturbed angles.
Raw marginal centers reproduce the same response; their largest coordinate difference from moment centers is below $5.0\times10^{-5}a$.
The weighted continuum first moment changes by at most $3.2\times10^{-4}a$, a diagnostic of lattice and full state corrections rather than an exact invariant of the lattice dynamics.
Initial and final fields, checkpoint diagnostics, independent center and orientation checks, and all control runs accompany the numerical evidence.

\FloatBarrier
\section{An independently prepared neutral \texorpdfstring{$\mathbb{CP}^{7}$}{CP7} moving pair}
\label{sec:cp7-travel}
A stretched pair can change shape and slow, so its transient center motion alone does not demonstrate translation of its whole profile.
Here we instead seek a field that solves the equation for translation at a prescribed velocity, and release that field without further forcing.
The full projector at every site is compared with a shifted copy of the launch profile.
The periodic full $\mathbb{CP}^{7}$ rest state is separately refined after collar removal.
Its energy is $0.9733858$ and maximum gradient $1.64\times10^{-11}$.
Translation curvatures are unresolved near zero; the six other computed low modes are positive.
This full stationary baseline differs from the common flag preparation used in Section~\ref{sec:matched-motion}.

For a prescribed direction $\bm e$, use the gauge covariant negative eighth order spatial derivative
\begin{equation}
 t_{\bm e}=-P^\perp\sum_{\mu=x,y}e_\mu\sum_{j=1}^4c_j
 \left(\psi_{i+j\hat\mu}^{\,\parallel i}-\psi_{i-j\hat\mu}^{\,\parallel i}\right) \,,
 \qquad(c_1,c_2,c_3,c_4)=\left(\frac45,-\frac15,\frac4{105},-\frac1{280}\right) \,.
 \label{eq:D8}
\end{equation}
Here, $\psi_j^{\,\parallel i}=\psi_j\langle\psi_i|\psi_j\rangle^*/|\langle\psi_i|\psi_j\rangle|$ aligns a neighbor ray to the central phase.
The direction is perpendicular to the pair's separation.
The first order equation $\mathcal H\chi=2\ii t_{\bm e}$ is solved after two Cartan phases and two approximate translation conditions are projected.
It gives $\operatorname{Re}\langle2\ii t_{\bm e},\chi\rangle=1.118\times10^5$, with projected relative residual $2.05\times10^{-11}$ and full relative residual $1.74\times10^{-7}$.
The maximum local $|\chi|$ is $2.139\times10^3$.
A finite amplitude first order guess is inadequate: even an orbit/normal exponential at $v=2\times10^{-5}$ has relative TDVP/translation mismatch $6.75$.
It is not used as a moving state.

The nonlinear corrector instead solves
\begin{equation}
 {\cal R}_v=\nabla E-2\ii v\,t_{\bm e}=0 \,,\qquad
 \dd{\cal R}_v=\mathcal H-2\ii v\,\dd t_{\bm e} \,,
 \label{eq:travel-residual}
\end{equation}
including the derivatives of neighbor phase alignment and the horizontal projector.
Generic finite difference checks validate this derivative in both local dimensions at approximately $10^{-11}$ relative error.
Phase and position conditions regularize Newton solves, but acceptance requires the complete physical residual to decrease.
The resulting moving profiles are listed in Table~\ref{tab:cp7-branch}.
The residual ratio is $\|{\cal R}_v\|/(2|v|\|t_{\bm e}\|)$.

\begin{table}[ht]
\caption{Nonlinear periodic $\mathbb{CP}^{7}$ moving preparations. Charges vary between preparations; this is not a dispersion relation at fixed Noether charges.}
\label{tab:cp7-branch}
\begin{ruledtabular}
\begin{tabular}{crrr}
$v$ & Energy & $E(v)-E(0)$ & Complete relative residual\\
$2\times10^{-5}$&0.9734049&$1.912\times10^{-5}$&$1.75\times10^{-7}$\\
$5\times10^{-5}$&0.9734859&$1.001\times10^{-4}$&$1.76\times10^{-7}$\\
$1\times10^{-4}$&0.9737095&$3.237\times10^{-4}$&$1.78\times10^{-7}$\\
\end{tabular}
\end{ruledtabular}
\end{table}
\FloatBarrier
At the lowest and highest velocities, the full residual norms are $1.87\times10^{-11}$ and $9.52\times10^{-11}$, versus projected norms $4.45\times10^{-13}$ and $8.23\times10^{-13}$.
Removed translation components have magnitudes $1.32\times10^{-11}$ and $6.7\times10^{-11}$; phase components remain below $3\times10^{-16}$.
Thus, the reported mismatch includes the reactions of the regularizing conditions.
An approximate D8 derivative is not an exact continuous lattice translation generator.

The branch is nonlinear even at these small velocities.
The coefficient $2[E(v)-E(0)]/v^2$ decreases from $9.560\times10^4$ to $6.475\times10^4$, compared with the linear response value $1.118\times10^5$.
The constituent cloud separation decreases from $12.28$ at rest to $11.89$, $11.43$, and $10.85$.
Vacuum subtracted Cartan sums change from approximately $(-112.1,-81.17)$ at rest to $(-118.0,-82.24)$ at the highest velocity.
Each physical release conserves its own launch charges; acceleration along this branch at fixed charges has not been constructed.

Table~\ref{tab:cp7-release} reports the full projector checks for releases at the lowest and highest velocities under the full periodic TDVP through $\tau=1000$.
The profile test uses the complete local $8\times8$ ray projector $P=\psi\psi^\dagger$.
It compares $P(\tau)$ with the prescribed Fourier translation of $P(0)$ by $v\bm e\tau$, without fitting displacement or Cartan phases.
The relative profile error is
\begin{equation}
 \epsilon_{\rm prof}(\tau)=
 \frac{\|P(\tau)-{\cal T}_{v\bm e\tau}P(0)\|_{\ell^2,F}}
 {\|P(0)-P_{\rm vac}\|_{\ell^2,F}} \,.
 \label{eq:profile-error}
\end{equation}
A separate circular center derived from $\langle\bar Q\rangle$ fits the velocity; it is not used to improve the profile prediction.
Fourier interpolation need not preserve projector idempotency exactly; the observed maximum errors are $1.8\times10^{-9}$ and $9.2\times10^{-9}$ and are included in the comparison.

\begin{table}[ht]
\caption{Independent full projector checks of the $\mathbb{CP}^{7}$ releases through $\tau=1000$. The highest velocity is run at both displayed steps.}
\label{tab:cp7-release}
\begin{ruledtabular}
\begin{tabular}{ccrrr}
$v$ & Step & Max $\epsilon_{\rm prof}$ & Max local projector error & $(v_{\rm fit}-v)/v$\\
$2\times10^{-5}$&0.1&$2.024\times10^{-9}$&$4.267\times10^{-9}$&$-4.64\times10^{-7}$\\
$1\times10^{-4}$&0.2&$1.103\times10^{-8}$&$2.512\times10^{-8}$&$-5.10\times10^{-7}$\\
$1\times10^{-4}$&0.1&$1.103\times10^{-8}$&$2.512\times10^{-8}$&$-5.10\times10^{-7}$\\
\end{tabular}
\end{ruledtabular}
\end{table}
At $v=10^{-4}$, step halving changes the full projector $\ell^2$ norm by at most $6.34\times10^{-11}$, the maximum site projector by $1.09\times10^{-11}$, and the fitted center by $4.11\times10^{-14}$.
Across these releases, energy changes are below $1.72\times10^{-12}$, local norm squared errors below $3.25\times10^{-12}$, and raw Cartan changes below $6.3\times10^{-11}$.
Ambient degree zero and both line degrees $-1$ are preserved to roundoff; principal gaps remain above $1-8.0\times10^{-8}$.
The fitted velocities agree with the prescribed values to $5.2\times10^{-7}$ relatively.
Actual travel is only $0.02$ and $0.1$ lattice units, much smaller than the molecule.

\section{Neutral \texorpdfstring{$\mathbb{CP}^{14}$}{CP14} trimer inertia and conservative motion}
\label{sec:trimer-motion}\label{sec:motion}
The neutral trimer provides a second composition for which a translating profile can be prepared and tested.
Its response to a small velocity also gives an inertia tensor: the energy cost differs for motion along and across the molecule.
We distinguish that linear response at fixed charges from the nonlinear moving preparations, whose charges can differ from one another.
The neutral trimer belongs to the same $(1,2)$ member of Equation~\eqref{eq:latticeH}.
Its periodic moving profiles have $\bm k=(-2,+1)$ and use the same full product dynamics as the pair.
With $J=-\ii$ on horizontal tangents, a translation source is $b_a=-2Jt_a$ and the internal response solves $\mathcal H\chi_a=b_a$.
Translations and exact Cartan phase modes are removed; linear charge constraints are imposed by a bordered correction.
If $X=\mathcal H^{-1}\mathsf B$ and $Y=\mathcal H^{-1}\mathsf C_Q$ for source columns $\mathsf B$ and charge covectors $\mathsf C_Q$, the fixed charge response is
\begin{equation}
 X_c=X-Y(\mathsf C_Q^TY)^{-1}\mathsf C_Q^TX \,,\qquad
 M_{ab}=\operatorname{Re}\langle b_a,\chi_b\rangle \,.
 \label{eq:chargecorrection}
\end{equation}
This is the symmetry reduced internal response relevant to relative equilibrium analysis~\cite{Simo}; including residual scale translation eigenvalues in a raw inverse would give a stencil dependent quantity.

At the $N=97$ collar reference the D8 response gives
$M_\parallel\simeq1.635\times10^5$ and $M_\perp\simeq3.250\times10^4$, along $(1,1)/\sqrt2$ and $(1,-1)/\sqrt2$.
Units are $\hbar^2/(E_0a^2)$.
Changing the spatial stencil from fourth to eighth order changes the matrix by $0.0951\%$; equation residuals are below $7.9\times10^{-12}$ relatively.
The explicit charge constraint changes the inertia matrix by only $3.19\times10^{-15}$ relatively and reduces first order charge residuals below $3.6\times10^{-10}$.
This near equality of constrained and unconstrained linear response does not make the nonlinear velocity branch charge independent.
The bordered construction establishes a linear response at fixed charges; it is not a nonlinear branch at fixed charges.
Both the pair and trimer finite velocity families used here contain preparations with different Cartan sums, each conserved separately during its own release.

Periodic correction of Equation~\eqref{eq:travel-residual} gives relative launch mismatch $2.29\times10^{-6}$ at $v=10^{-5},10^{-4},2\times10^{-4}$ on $N=97$, and at $v=2\times10^{-4}$ on $N=129$.
At $v=10^{-5}$, $2\Delta E/v^2\simeq3.250\times10^4$, exceeding the small velocity inertia by only $5.3\times10^{-5}$ relatively.
Cartan sums differ between branch points: changes from rest at $v=10^{-5},10^{-4},2\times10^{-4}$ are respectively
$(-0.008020,0.0004338)$, $(-0.8176,0.04478)$, and $(-3.492,0.1991)$.
Each release conserves the charges of its own preparation.
At larger trial speeds, continuation reaches $v=3.5\times10^{-4}$ and $3.75\times10^{-4}$ with complete relative residuals $2.28\times10^{-6}$ and $2.27\times10^{-6}$.
Bounded attempts at $4\times10^{-4}$ and $5\times10^{-4}$ stall at $0.0408$ and $0.217$; these iterates are not accepted moving preparations or evidence of a limiting speed.
No long trajectory from the faster preparations is reported.
The propagation tests below cover a distance $\sqrt2a$.

\subsection{Conservative projective releases}
\begin{figure}[t]
\includegraphics[width=\textwidth]{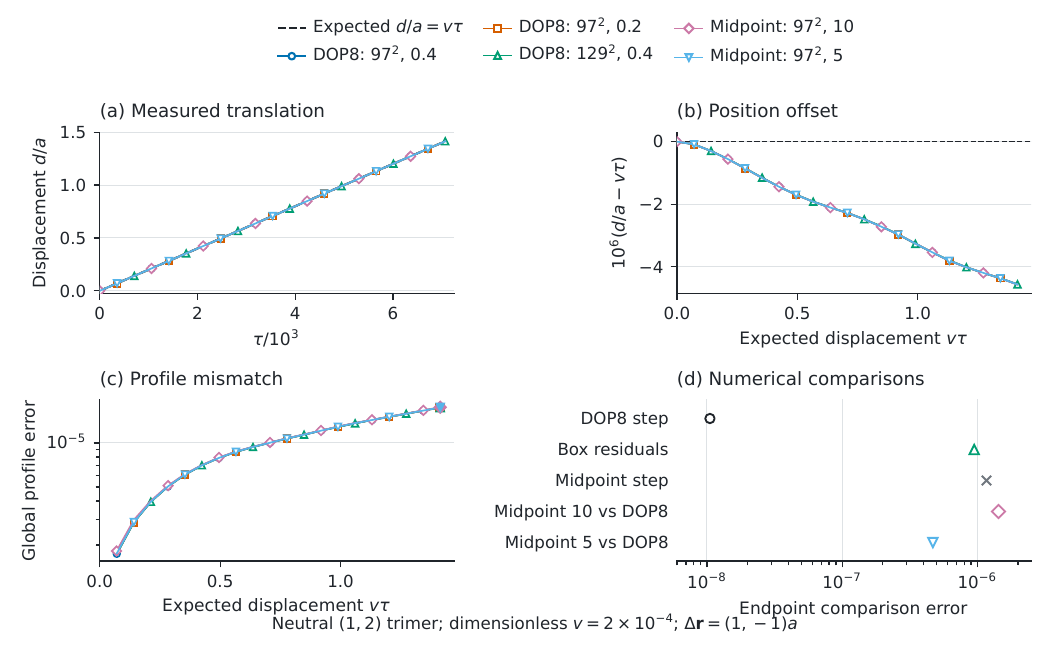}
\caption{Finite motion with numerical controls.
(a) Displacement $d/a$ along $(1,-1)/\sqrt2$ and (b) its deviation from $v\tau$, with dimensionless speed $v=2\times10^{-4}$ and time $\tau=E_0t/\hbar$.
(c) Full projector mismatch to the translated launch field; fractional shifts use Fourier translation of the projector, while filled endpoints use the exact integer shift.
(d) Endpoint step, box, and integration method comparisons; the midpoint endpoints are compared to the finer DOP853 endpoint.
The box comparison is the difference of translation residual fields, not the raw cross box field difference.
The physical speed is $vE_0a/\hbar$; the observed displacement $\sqrt2a$ is smaller than the composite diameter.
}
\label{fig:motion}
\end{figure}

The physical releases integrate Equation~\eqref{eq:TDVP}.
All five runs in Table~\ref{tab:motion} start from the corrected $v=2\times10^{-4}$ state and reach $T=\sqrt2/v$.
Figure~\ref{fig:motion} compares their motion and numerical controls.
The primary endpoint comparison uses its exact integer shift $(1,-1)$.
The distance between the evolved endpoint and the shifted launch ray is defined in Equation~\eqref{eq:distance}.
The Letter calls this $d_{\rm prof}$: for normalized projectors it equals $[\tfrac12\sum_i\|P_i(\tau)-P_i^{\rm shift}(0)\|_F^2]^{1/2}$, summed over all sites without division by lattice area.
In Figure~3 of the Letter, Method~1 is DOP853 with actual step approximately $0.2$, and Method~2 is implicit midpoint with actual step approximately $5$; Table~\ref{tab:motion} gives the interval counts and the other controls.
We evaluate the small difference with a cancellation resistant, locally phase aligned formula.
No spatial fit or global Cartan fit enters this comparison, and the distance is not divided by the number of sites.

To put the profile error on a physical scale, we compare the actual launch field with its exact periodic shift by $(1,-1)a$ in the same metric.
This gives $d_{\rm shift}\simeq5.26$, the error an unmoved launch profile would register against the prescribed endpoint target.
The finer DOP853 endpoint has $d_{\rm prof}\simeq1.75\times10^{-5}$, or $d_{\rm prof}/d_{\rm shift}=3.32\times10^{-6}$.
Direct projector differences, a stable phase aligned ray formula, and exterior products give consistent distances from the saved launch and endpoint fields.
This reference contextualizes the mismatch without changing its normalization or fitting the trajectory.

\begin{table}[!hb]
\caption{Neutral $\mathbb{CP}^{14}$ trimer: completed conservative releases at the same prescribed endpoint time.
The last columns give the distance to the exact integer shifted launch field and the largest local ray error.
The displayed steps are rounded; the exact steps are $T/n_t$, with $n_t=17678$, $35356$, $17678$, $708$, and $1415$, respectively.}
\label{tab:motion}
\begin{ruledtabular}
\begin{tabular}{llrrr}
Method & Box & $\Delta\tau$ & $d$ & Maximum local error\\
DOP853 & $97^2$ & $0.4000$ & $1.746\times10^{-5}$ & $1.269\times10^{-6}$\\
DOP853 & $97^2$ & $0.2000$ & $1.746\times10^{-5}$ & $1.269\times10^{-6}$\\
DOP853 & $129^2$ & $0.4000$ & $1.746\times10^{-5}$ & $1.268\times10^{-6}$\\
Implicit midpoint & $97^2$ & $9.987$ & $1.754\times10^{-5}$ & $1.270\times10^{-6}$\\
Implicit midpoint & $97^2$ & $4.997$ & $1.747\times10^{-5}$ & $1.270\times10^{-6}$\\
\end{tabular}
\end{ruledtabular}
\end{table}

\begin{samepage}
The finer DOP853 and midpoint endpoints differ by $4.69\times10^{-7}$ in the same full ray distance.
The DOP853 step comparison is $1.05\times10^{-8}$; the midpoint step comparison is $1.17\times10^{-6}$.
At the same DOP853 step, increasing the box from $N=97$ to $N=129$ changes the endpoint error magnitude $d$ by $2.57\times10^{-10}$. The central translation residual fields differ by $9.49\times10^{-7}$.
This compares changes in the fields rather than their raw finite box difference.
Across saved checkpoints the maximum local norm error is $2.64\times10^{-12}$, normalized ray energy change $2.73\times10^{-12}$, and absolute Cartan change $8.74\times10^{-11}$.
No normalization or charge correction is applied during evolution.
The charge pattern persists; reported DOP853 constituent separations and sizes change by less than $10^{-7}$ after the translation.
\par
\end{samepage}

The independent midpoint control is needed because DOP853 at these steps attenuates hypothetical fast normal oscillations over the full interval.
Midpoint uses the canonical extension of the normalized ray energy to unrestricted complex stage coordinates.
If $n_i=\psi_i^\dagger\psi_i$, its off unit horizontal right hand side is divided by $n_i$; omitting this factor would violate the Cartan conservation test on generic unequal norm fields.
All actual implicit stage residuals are below $3\times10^{-13}$.
Generic full space tests verify the dense right hand side, second order convergence, and reversibility.
Exact midpoint preserves quadratic invariants and has unit linear oscillator amplitude, but its large time steps distort fast phases.
The method and step comparisons support this prepared slow trajectory, not arbitrary perturbed state stability or fast mode spectroscopy.
The endpoint relative D8 velocity mismatch grows to $1.74\times10^{-4}$ for finer DOP853 and $7.13\times10^{-4}$ for finer midpoint.
Neither exact rigid propagation nor asymptotic absence of radiation is established.

\FloatBarrier
\section{Translation reactions in the charged partners}
\label{sec:charged-corrector}
A charged molecule should retain a translation reaction in the conservative continuum, even when its internal shape is adjusted.
Applying the same moving profile calculation to the charged partners tests this prediction and exposes a reaction that a constrained solver could otherwise hide.
The continuum identity below, rather than nonconvergence alone, supplies the physical interpretation.
We apply the same nonlinear corrector to the periodic charged $(1,2)$ pair and $(1,1)$ trimer at $v=10^{-4}$ along $(-1,1)/\sqrt2$.
Before correction, the rest fields are independently refined to maximum gradients $3.9\times10^{-12}$ and $4.7\times10^{-12}$.
Both bounded attempts reduce the complete residual ratio from approximately unity, but retain the translation reaction shown in Table~\ref{tab:charged-corrector}.
Projecting this reaction out would incorrectly accept a forced state as an autonomous translator.

Let $t_x,t_y$ be the D8 translation tangents and define their symplectic pairing by
\begin{equation}
 S_{xy}=2\operatorname{Im}\langle t_x,t_y\rangle \,.
\end{equation}
In the translation invariant continuum, $\operatorname{Re}\langle t_a,\nabla E\rangle=0$; hence Equation~\eqref{eq:travel-residual} implies
\begin{equation}
 \operatorname{Re}\langle t_x,{\cal R}_{\bm v}\rangle=v_y S_{xy} \,,\qquad
 \operatorname{Re}\langle t_y,{\cal R}_{\bm v}\rangle=-v_x S_{xy} \,.
 \label{eq:charged-reaction}
\end{equation}
The common representation weighted Berry degree therefore requires $C_{\rm full}\bm v=0$ for an isolated rigid continuum translator.
For the lattice fields, we directly evaluate both the energy term and the full residual rather than assume exact continuous translation symmetry.
The measured pairings $S_{xy}=6.283$ and $-6.283$ are close to $2\pi C_{\rm full}$.
Each Cartesian residual pairing is $4.443\times10^{-4}$ for the pair and $-4.443\times10^{-4}$ for the trimer.
For projection onto the two translation tangents, the ratio of projected to total residual norms exceeds $1-2\times10^{-6}$ in both cases.

\begin{table}[ht]
\caption{Complete translating equation residuals at $v=10^{-4}$. Neutral entries are accepted moving preparations; charged entries are final iterates of bounded attempts with eight steps and are not moving solutions.}
\label{tab:charged-corrector}
\begin{ruledtabular}
\begin{tabular}{lcr}
Composition and representation & $C_{\rm full}$ & $\|{\cal R}_v\|/(2|v|\|t_{\bm e}\|)$\\
Pair $(1,1)$ & 0 & $1.78\times10^{-7}$\\
Pair $(1,2)$ & $+1$ & $0.2914$\\
Trimer $(1,1)$ & $-1$ & $0.2974$\\
Trimer $(1,2)$ & 0 & $2.29\times10^{-6}$\\
\end{tabular}
\end{ruledtabular}
\end{table}
These calculations identify the reaction retained by the charged attempts and corroborate the continuum obstruction.
Failure of a bounded nonlinear solve alone is not a proof that a lattice relative equilibrium cannot exist.
The result does not prohibit externally driven charged motion, turning trajectories, or nonrigid propagation.

\section{Design parameters and scope of a local occupation encoding}
\label{sec:encoding}\label{sec:exchange-weight-origin}
The eight and fifteen state multiplets can be built algebraically from fermion orbitals with three internal colors.
One fermion carries the fundamental $\bm3$; two fermions carry its conjugate $\bar{\bm3}$, labeled by the missing color.
The constructions below show how the local multiplet and attraction can be encoded.
They do not derive an experimental implementation of the complete exchange pattern and anisotropy in Equation~\eqref{eq:latticeH}.
The positive weights in Equation~\eqref{eq:parameters} are independent design parameters, with $w_B/w_F\simeq1.693$.
Together with the displayed lattice coefficients, they specify the effective Hamiltonian completely.
At $\alpha/\rho=20$, the measured constituent radii are about $5$--$8$ lattice spacings, so the reference textures are spatially resolved.
Neither this ratio nor the weights are universal or fitted to a material. The additional recorded digits specify the numerical design point and are retained only for reproducibility.

For the eight state adjoint member $(1,1)$, two orbitals suffice.
Two distinguishable three color fermion orbitals, $g,e$, with occupations $(1,2)$ carry $\bm3\otimes\bar{\bm3}=\bm8\oplus\bm1$.
The quadratic Casimir has eigenvalues $3$ and $0$, so
\begin{equation}
 P_8=\frac{T^aT^a}{3} \,,\qquad
 H_{\rm H}=J_{\rm H}(3I-T^aT^a) \,,\qquad
 K_9=(1-n_{g,1})n_{e,3} \,.
 \label{eq:adjoint-encoding}
\end{equation}
For $J_{\rm H}>0$ the selector separates the singlet by $3J_{\rm H}$; projecting $K_9$ gives precisely $K=U_{1,1}^\dagger K_9U_{1,1}$.
In the hole color basis $n_{e,3}=\operatorname{diag}(1,1,0)$, which verifies the attraction directly.
The bare density coupling can mix singlet and octet, so exact encoding requires $P_8K_9P_8$; a finite selector gap introduces virtual corrections.
This eight state multiplet is not an ordinary spin-$1$ local moment, which has only three states.
This realizes the local $SU(3)$ representation algebraically; the exchange between such sites remains to be engineered.

The fifteen state member uses one additional orbital. Its occupation encoding also gives a direct interpretation of the onsite attraction.
Take three distinguishable fermionic orbitals, each with three colors, and impose occupations $(N_0,N_1,N_2)=(1,2,2)$.
The one particle orbital carries $\bm3$ and each two particle orbital carries $\wedge^2\bm3=\bar{\bm3}$ in its hole color basis.
The decomposition and an irreducible representation selector are
\begin{equation}
 \bm3\otimes\bar{\bm3}\otimes\bar{\bm3}
 =(1,2)\oplus(2,0)\oplus2(0,1) \,,\qquad
 H_{\rm H}=J_{\rm H}\left(\frac{16}{3}-T^aT^a\right) \,,\qquad
 P_{15}=\frac18\left(T^aT^a-\frac{10}{3}\right)\left(T^aT^a-\frac43\right) \,.
 \label{eq:encoding}
\end{equation}
The selector has gaps $2J_{\rm H}$ and $4J_{\rm H}$ to the other sectors.
It selects the irreducible representation, not the nonlinear flag manifold.
Writing $n_{o,\alpha}=c^\dagger_{o,\alpha}c_{o,\alpha}$, the desired preprojection density operator is
\begin{equation}
 K_{27}=\frac12(1-n_{0,1})(n_{1,3}+n_{2,3}) \,,\qquad K=U_{1,2}^\dagger K_{27}U_{1,2} \,.
 \label{eq:K-density-encoding}
\end{equation}
This uses $\mathsf A=1-n_{0,1}$ and, in a two particle hole basis, $\mathsf C=n_{o,3}$.
The bare density operator does not preserve the irreducible representation.
An exact encoded interaction uses $P_{15}K_{27}P_{15}$; a finite Hund gap gives this projection only to leading order, with virtual corrections requiring separate control.
The projection defines the desired effective local interaction; implementing it together with the full lattice exchange remains a separate physical problem.

For implementation entirely in $V_{1,2}$, an equivalent operator is
\begin{equation}
 K=\frac7{18}I-\frac56F^3+\frac{F^8}{6\sqrt3}
 -\frac{2T^8}{3\sqrt3}
 +\frac{F^3T^8}{\sqrt3}+\frac{F^8T^8}{3} \,.
 \label{eq:K-polynomial}
\end{equation}
The displayed factors commute with $T^8$.
This follows by substituting $\mathsf A=2I/3-t^3-t^8/\sqrt3$ and
$\overline{\mathsf C}=2I/3-(\bar t_1^8+\bar t_2^8)/\sqrt3$ before compression, then using total $T^8$ to reduce the fundamental products.
It does not replace compression of a product by a product of compressions.

\section{Evidence, reproducibility, and limits}
\label{sec:data}
The numerical evidence supports three distinct conclusions: local survival after all local amplitudes in the product ansatz are relaxed, energy ordering against specified breakup fields, and finite time motion of specified preparations.
The exact analytical results establish the common flag energy, the topology map, and the quantum vacuum bounds for the stated Hamiltonian and assumptions.
They do not by themselves establish a texture minimum.

\subsection{Independent checks and supplied evidence}
The accompanying evidence, summarized in Table~\ref{tab:evidence}, contains the four reference fields on $N=97$, their physical operators, and all 32 reported low eigenvectors.
Each field is paired with the spectrum evaluated at that field.
An independent program using NumPy reconstructs constituent operators, bond energies, horizontal gradients, charges, tangent coordinates, and full Hessian actions, without importing the original optimizer or eigensolver.
All 80 checks pass after extraction into a separate environment; the largest unprojected residual among the 32 eigenpairs is $1.85\times10^{-12}$.
Centered energy differences test the gradient away from stationarity, and centered gradient differences test the full Hessian.
Self-adjointness in the real tangent metric, covariance under phase changes, mode normalization, and symmetry overlaps provide additional checks.
Translations remain in these spectra; only the two exact global Cartan phase directions are removed.

\begin{table}[ht]
\caption{Scientific data supplied with the calculations. The companion archive contains final numerical records and programs for verifying saved fields and regenerating figures.}
\label{tab:evidence}
\begin{ruledtabular}
\begin{tabular}{p{0.30\textwidth}p{0.62\textwidth}}
Evidence category & Contents and purpose\\
Static molecules & Four full states, local operators, 32 low eigenvectors, and independent energy, gradient, charge, and Hessian checks\\
Binding and parameter changes & Attraction and stiffness scans, separately optimized fragments, and larger lattice comparisons\\
Escape paths & Explicit joined image fields, bounds on the energy between images, and checks of spatial regularity; the unsuccessful regular fission construction is distinguished\\
Matched dynamics & Pair and trimer releases from common flag preparations, corresponding flag evolution, center and orientation measurements, time step controls, and unperturbed releases\\
Prepared motion & Neutral translating profiles, full projective evolution, comparison with prescribed translations, integration controls, and the unsuccessful charged and faster profile solves\\
Figures & Plotted data and scripts that regenerate the figures\\
\end{tabular}
\end{ruledtabular}
\end{table}
\FloatBarrier

The companion archive \texttt{numerical\_data.zip}, deposited at Zenodo (\href{https://doi.org/10.5281/zenodo.22959390}{doi:10.5281/zenodo.22959390}), groups the final data by calculation and provides verification programs, figure scripts, and checksums.
Its \texttt{DATA\_FORMATS.md} describes the array axes, basis conventions, normalization, units, and boundary conditions; \texttt{ARRAY\_INDEX.json} lists the stored arrays.
The portable static and path checks use Python and NumPy; the figure scripts also require Matplotlib.
These programs verify saved results and regenerate figures; they do not rerun the complete minimizations, continuation, or time evolution.
Verification and figure regeneration were tested with Python 3.12.14, NumPy 2.3.5, and Matplotlib 3.10.8.
SciPy and JAX are not required by the supplied programs; the original simulation environment additionally used SciPy 1.17.0 and JAX 0.9.0.1.
Independent C++ dynamics implementations use the same operators in a common Cartan basis and are checked against dense gradients.
The archive does not include the optimization, continuation, or integration drivers, their C++ source programs, eigenvectors beyond the central 32 modes, intermediate trajectory wavefunctions, the complete set of flag relaxation fields, or the unsuccessful fission path image arrays. For these calculations it supplies the final numerical records and the explicitly listed fields and controls.

\subsection{What the results establish}
Positive computed low curvatures and independent searches provide numerical evidence for local minima, rather than a rigorous certificate that every eigenvalue is nonnegative.
In particular, a positive bound for the normal principal block does not establish positivity of its coupled Schur complement.
The neutral $(1,2)$ trimer survives the sampled attraction window and separate stiffness and box controls.
A finite grid cannot exclude a crossing between samples or locate an upper stability edge; the other composition--representation combinations do not have the same parameter survey.
Establishing a rigorous interval would require validated stationary solutions, full Hessian lower bounds, and binding bounds between samples.

The binding comparisons use specified stationary fragments.
They neither prove the global infimum in a fragment sector nor exhaust partitions with other winding signs, and the compared fields are not required to have matching Cartan charges.
A charge constraint can raise a global fragment infimum, but the measured relaxed fields are not certified infima, so that observation gives no constrained decay threshold here.
The continuous vacuum contraction gives a numerical upper bound on an unconstrained lattice escape barrier.
It gives no positive lower bound or minimum saddle, and its changing Cartan sums exclude its interpretation as a decay trajectory in a closed system.
The bound concerns the lattice Hamiltonian, even for the path with a regular spatial interpolation.
A regular fission path remains uncomputed.

The matched releases test small, finite time responses to specified deformations; no interval in box size or preparation amplitude is established for them.
The ratio of species displacements supports the prediction from the Berry weights, while the slowing already present in flag dynamics has no identified mechanism here.
The separately prepared neutral profiles retain their shape over the stated short distances.
The spectrum about the moving state, the stability of its fast modes, long distance transport, and asymptotic radiation remain to be studied.
Charged and faster profile solves that do not meet the complete residual criterion remain unsuccessful calculations, rather than evidence for a general nonexistence theorem.

Finally, these textures are mean field excitations: unrestricted refers to local amplitudes within a product of site states.
Quantum tunneling, entangled soliton lifetimes, and thermal lifetimes are not calculated.
The general representation construction supplies candidate neutral compositions, not a bound molecule for every representation.
The specified weights, anisotropy, and exchange pattern are design choices, not universal values or a microscopic derivation for a material.
Implementing simpler interactions would require a new texture stability calculation.